\documentclass{aa}  

\usepackage[utf8]{inputenc}

\usepackage{amsmath,amsfonts,amssymb}
\usepackage{upgreek}
\usepackage{graphicx}
\usepackage{txfonts}
\usepackage[colorlinks=true, allcolors=blue]{hyperref}
\usepackage{natbib}
\usepackage{widetext}
\usepackage{svg}
\usepackage{textgreek}

\newcommand{\V}[1]{\boldsymbol{#1}} 
\newcommand{\Vx}{\V{x}} 

\newcommand{\GP}[1]{\left(#1\right)}
\newcommand{\GC}[1]{\left[#1\right]}

\newcommand{\EST}[1]{\widehat{#1}}

\usepackage{xcolor}
\newcommand{\commentdaniel}[1]{\textcolor{black}{#1}} 

\newcommand{\correction}[1]{\textcolor{black}{#1}} 

\newcommand{\commentelsa}[1]{\textcolor{black}{#1}}

\usepackage{ulem}

\newcommand{\revision}[1]{\textcolor{black}{#1}} 

\newcommand{\rev}[1]{\textcolor{black}{#1}}

\title{Single-aperture spectro-interferometry in the visible at the Subaru telescope with FIRST: \rev{F}irst on-sky demonstration on Keho`oea~($\alpha$~Lyrae)~and~Hokulei~($\alpha$~Aurigae)}

\titlerunning{FIRST at the Subaru Telescope}

\author{%
S.~Vievard \inst{\ref{a},\ref{b}} \and %
E.~Huby\inst{\ref{c}} \and %
S.~Lacour\inst{\ref{c}} \and %
O.~Guyon\inst{\ref{a},\ref{b},\ref{d1},\ref{d2}} \and %
N.~Cvetojevic \inst{\ref{e}} \and %
N.~Jovanovic\inst{\ref{f}} \and %
J.~Lozi\inst{\ref{a}} \and %
K.~Barjot\inst{\ref{c}} \and %
V.~Deo\inst{\ref{a}} \and %
G.~Duch\^ene\inst{\ref{g},\ref{h}} \and %
T.~Kotani \inst{\ref{b},\ref{i},\ref{j}} \and %
F.~Marchis\inst{\ref{k}} \and %
D.~Rouan\inst{\ref{c}} \and %
G.~Martin\inst{\ref{h}} \and %
M.~Lallement\inst{\ref{c}}  \and %
V.~Lapeyrere\inst{\ref{c}} \and %
F.~Martinache\inst{\ref{e}}  \and %
K.~Ahn \inst{\ref{a}} \and %
N.~Skaf \inst{\ref{a},\ref{c}}\and %
M.~Tamura\inst{\ref{b},\ref{j},\ref{l}} \and %
D.~Leilehua~Yuen\inst{\ref{m}} \and %
A.~Leinani~Lozi\inst{\ref{n}} \and %
G.~Perrin\inst{\ref{c}}}

\institute{National Astronomical Observatory of Japan, Subaru Telescope, 650 North Aohoku Place, Hilo, HI 96720, U.S.A. email: vievard@naoj.org \label{a}
\and Astrobiology Center, 2-21-1, Osawa, Mitaka, Tokyo, 181-8588, Japan \label{b} 
\and LESIA, Observatoire de Paris, 5 pl. Jules Janssen, 92195 Meudon, France \label{c}
\and Steward Observatory, University of Arizona, Tucson, AZ 85721, USA \label{d1}
\and College of Optical Sciences, University of Arizona, Tucson, AZ 85721, U.S.A. \label{d2} 
\and Observatoire de la Côte d'Azur, 96 Boulevard de l'Observatoire, 06300 Nice, France \label{e}
\and California Institute of Technology, 1200 E California Blvd, Pasadena, CA 91125, U.S.A. \label{f} 
\and Astronomy Department, University of California, Berkeley, CA 94720-3411, USA \label{g}
\and Univ. Grenoble Alpes, CNRS, IPAG, 38000 Grenoble, France \label{h}
\and Department of Astronomical Science, The Graduate Un iversity for Advanced Studies, SOKENDAI, 2-21-1 Osawa, Mitaka, Tokyo 181-8588, Japan \label{i}
\and National Astronomical Observatory of Japan, 2-21-1 Osawa, Mitaka, Tokyo 181-8588, Japan \label{j}
\and Carl Sagan Center at the SETI Institute, 189 Bernardo Av., Mountain View, CA 94043, USA \label{k} 
\and Department of Astronomy, Graduate School of Science, The University of Tokyo, Tokyo, Japan \label{l}
\and Hawaiian cultural practitioner, Hālau Lei Manu, Hāla`i, Hilo, Hawai`i \label{m}
\and TMT International Observatory, 111 Nowelo St. Hilo, HI 96720 \label{n}}

\abstract{}
{FIRST is a spectro-interferometer combining\rev{,} in the visible\rev{,} the techniques of aperture masking and spatial filtering thanks to single-mode fibers. By turning a monolithic telescope into an interferometer, this instrument aims \rev{to deliver} high contrast capabilities at spatial resolutions that are inaccessible to classical coronagraphic instruments.}
{The technique implemented in the FIRST instrument is called pupil remapping: the telescope pupil is divided into \rev{subpupils} by a segmented deformable mirror conjugated to a micro-lens array injecting light into single-mode fibers. The fiber outputs are \rev{rearranged} in a \rev{nonredundant} configuration, allowing simultaneous measurement of all baseline fringe patterns. The fringes are also spectrally dispersed, increasing the coherence length and providing precious spectral information. The optical setup of the instrument has been adapted to fit onto the SCExAO platform at the Subaru Telescope.}
{We present the first on-sky demonstration of the FIRST instrument at the Subaru telescope. We \rev{used} eight \rev{subapertures} of the $8.2$-meter diameter pupil, each with a diameter of about $1$~m. Closure phase measurements \rev{were} extracted from the interference pattern to provide spatial information \rev{on} the target. We tested the instrument on two types of targets~: a point source (Keho`oea - $\alpha$~Lyrae, $m_R$~=~0.1) and a binary system (Hokulei - $\alpha$~Aurigae, $m_R$~=~-0.52, \rev{and a semi-major} axis = 56.4~mas). An average accuracy of $0.6^\circ$ is achieved on the closure phase measurements of Keho`oea, with a \revision{statistical error of about $0.15^\circ$ at best. We estimate that the instrument can be sensitive to structures down to a quarter of the telescope spatial resolution.} We \rev{measured} the relative positions of Hokulei Aa and Ab with an accuracy $\lesssim 1$~mas. }
{FIRST opens new observing capabilities in the visible wavelength range at the Subaru Telescope. \rev{With} SCExAO being a testing platform for high contrast imaging instrumentation for future 30-meter class telescopes, the successful demonstration and exploitation of FIRST is an important stepping stone for future interferometric instrumentation on extremely large telescopes.}{}

\begin{document} 
\maketitle

\section{Introduction}
\label{sec:intro} 

One of the key challenges in ground-based astronomy is the study of circumstellar environments, \rev{for example} the detection and characterization of stellar companions \rev{such as} exoplanets or protoplanets. This is achieved by using techniques qualified either as indirect (\rev{for example} radial velocity \rev{and} transit) or direct. The latter usually consists in masking the light of a star in the focal plane, using a coronagraph, to reveal its circumstellar environment. One challenge is to correct for atmospheric turbulence\rev{ as best as possible}. The latter induces wavefront errors that cause stellar leakage in coronagraphs, which limits the performance of detection in terms of \rev{the} inner working angle and contrast. By using \rev{a}daptive \rev{o}ptics \rev{(AO)} \citep{rousset1990first}\rev{,} and even extreme AO (ExAO)\rev{,} a large fraction of the atmospheric turbulence is corrected, and therefore \rev{it recovers} the telescope native resolution power, almost down to the diffraction limit. This is critical for confining most of the starlight behind the focal plane mask, \rev{and} thus increasing the achievable contrast.
Indeed, exoplanets are extremely faint compared to their host star by a factor of $10^{-6}$ and $10^{-9}$ for hot Jupiters and Earth-like planets\rev{,} respectively~\citep{seager2010exoplanet}. About 20 exoplanets have been imaged thanks to coronagraphs, but at distances relatively \rev{large} from their host star, \rev{that is} $\geq$ 10 astronomical units (au) \citep{Marois2008,Lagrange2010,Currie2014}. Closer-in planets, while more numerous, are harder to image due to the inner working angle (IWA) of coronagraphs being typically greater than $2\lambda/D$ with $\lambda$ \rev{being} the wavelength and $D$ the telescope diameter. For an 8-meter class telescope, this corresponds to about $70$~mas at $1550$~nm. 
Indirect detections \rev{have shown} that there is an abundance of massive planets around Sun-like stars at distances \rev{of} about 1 to 5 au \citep{pascucci2019exoplanet}. At the distance of the Taurus group, the closest stellar formation region from Earth (140~parsecs), this corresponds to an angular separation of $7$ to $40$~mas. These separations are then out of reach for current 8-meter telescopes using coronagraphs, \rev{which are} limited to regions >10 au for stars at about 140 pc.

One solution to probe this region is single telescope interferometry: sparse aperture masking \rev{(SAM)}\citep{haniff1987first}\rev{,} consists in placing a \rev{nonredundant} mask delimiting several \rev{subpupil}s in a pupil plane of the telescope and recombining the coherent light from all \rev{of} those \rev{subpupil}s. The \rev{nonredundancy} of the mask allows for each baseline (vector defined by each pair of \rev{subpupils}) to be unique. In the Fourier domain, \rev{this} means that the information carried by every fringe pattern is located at a unique spatial frequency and can thus be retrieved, independently from all other fringes. SAM has been regularly used on large telescope instruments including NACO\rev{/Very Large Telescope} \citep{lacour2011sparse,lagrange2012insight}, NIRC2/Keck \citep{hinkley2011observational}, VAMPIRES/Subaru~\citep{norris2015vampires}, SPHERE/\rev{Very Large Telescope} \citep{cheetham2016sparse}, or NIRISS/\rev{James Webb Space Telescope}~\citep{sivaramakrishnan2012non}. \rev{SAM is also planned} for future extremely large telescopes \citep[MICADO/\rev{European Extremely Large Telescope} - ][]{lacour2014aperture}.
The advantage of this technique is its sensitivity to structure\rev{s} that extend down to $0.5\lambda/B_{max}$ (IWA), with $B_{max}$ the interferometer\rev{'s} longest baseline \citep{lacour2011sparseB}. Considering $B_{max}$ as the diameter of the telescope, aperture masking beats the single telescope diffraction limit \rev{by a factor of two}, and brings the detection limit down to $10$~mas or less on an 8-meter telescope in the visible. Moreover, it allows \rev{one} to compute the closure phase (CP) self-calibrated quantity, which cancels out the AO residual and quasi-static aberrations in the instrument. The drawbacks of using aperture masking are \rev{the following: i)} it can only exploit a fraction of the pupil because of the \rev{nonredundancy} requirement, and \rev{ii)} remaining speckle noise over each \rev{subpupil} limits the achievable contrast to about~$10^{-3}$~\citep{gauchet2016sparse}.
A method that can make use of the aperture masking concept without its disadvantages is the pupil remapping technique. It consists in sampling the whole pupil and, with the help of single-mode fibers, recombining the different \rev{subpupil}s pairwise~\citep{perrin2006high}. This allows \rev{one to do the following: i)} exploit the light from the whole pupil, and \rev{ii)} "clean" the wavefront over each \rev{subpupil}\rev{, and therefore} remove the speckle noise\rev{,} thanks to the spatial filtering performed by the single-mode fibers\rev{;} and \rev{iii)} sense all spatial frequencies independently. \cite{lacour2007high} showed that such \rev{an} instrument could theoretically reach \commentelsa{a contrast of} $10^{-6}$ using a classical AO system. As shown in \cite{norris2014high}, the achievable contrast of pupil remapping instruments relies on the AO performance. Hence, recent advances in wavefront control~\citep{guyon2020validating} are expected to benefit such instruments.

An instrument with such features, named Fibered Imager foR a Single Telescope \rev{(FIRST)}, was developed at \rev{the} Paris Observatory \citep{kotani2009pupil}. FIRST is a fibered pupil remapper coupled with a spectrograph, operating in the \rev{v}isible ($600$ to $800$~nm range). 
FIRST demonstrated its capability to resolve the binary system Hokulei~\citep[$\alpha$ Aurigae or Capella, V= 0.08, R=-0.52 \rev{;} ][]{huby2013first} and perform spectroscopic analysis at the diffraction limit on the 3-meter Lick Telescope. This success led to moving the instrument on a larger telescope: it is currently installed on the $8.2$-meter Subaru Telescope as part of the Subaru Coronagraphic Extreme Adaptive Optics (SCExAO) instrument~\citep{2015PASP..127..890J,vievard2020capabilities}. The advantages of deploying FIRST behind an ExAO system with exquisite wavefront stability are twofold\revision{, and should significantly improve the overall performance of the instrument}. First, the coupling efficiency into the single-mode fibers is increased~\citep{jovanovic2017efficient}, hence optimizing the overall throughput. Second, interference fringes are stabilized, making long exposure times possible (up to almost one second) and increasing the instrument sensitivity to faint targets.
Having a visible wavelength, single aperture remapping interferometer behind an ExAO system on an $8.2$-m telescope uniquely positions FIRST to provide new observing capabilities. Specifically it could provide access to key spectral structures such as H\textalpha.

In Section~\ref{Sec:first-at-Subaru} we provide a comprehensive description and characterization of FIRST on SCExAO. As a first on-sky demonstration on the Subaru Telescope, we present the detection of the Hokulei~\footnote{\textit{This paper \rev{refers} to the main targets per their Hawaiian names, in honor of Mauna Kea, the mountain from which we were most fortunate to perform our observations, and with acknowledgment of the indigenous \rev{h}awaiian communities who have had and continue to have Kuleana (translated into English as "responsibility") to this land. We offer the reader some more information about these Hawaiian names in Appendix~\ref{Appendix-hawaiian-names}}} system in Section~\ref{Sec:first-on-sky}.

\begin{figure*}[h!]
    \centering
    \includegraphics[width=0.95\linewidth]{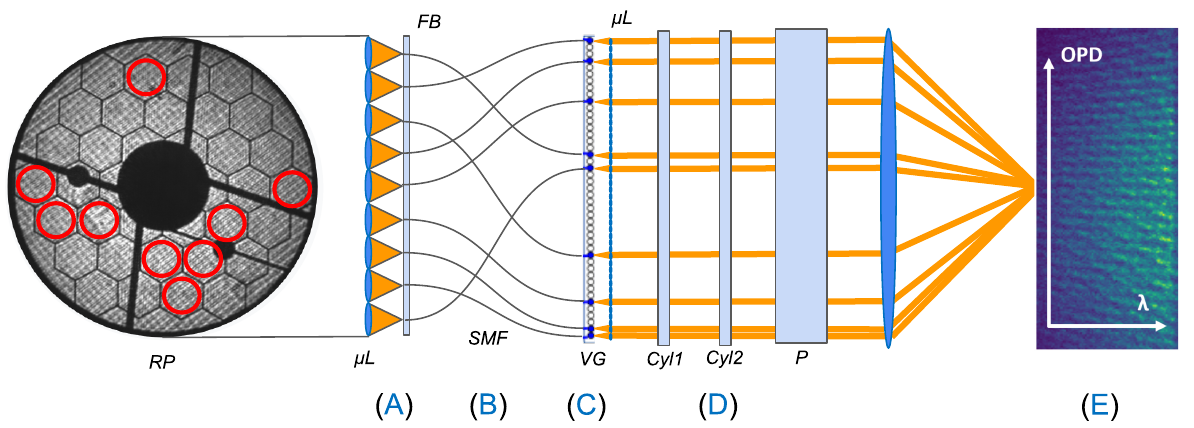}
    \caption{FIRST instrument principle: the redundant pupil of the telescope is divided into several \rev{subaperture}s (\textbf{A}). The selected \rev{subaperture}s, circled in red, are injected into single-mode fibers (\textbf{B}) and \rev{rearranged} into a \rev{nonredundant} pattern (\textbf{C}). The signal of each \rev{subaperture} is dispersed using anamorphic optics and a prism (\textbf{D}). Coherent recombination produces the interferometric signal on the camera, delivering Optical Path Difference (OPD) information as a function of wavelength (\textbf{E}). RP = Redundant Pupil - $\mu$L = micro-lenses - FB = Fiber Bundle - SMF = Single Mode Fibers - VG = V-groove - Cyl = Cylindrical lenses - P = Prism.}
    \label{fig:first-principle}
\end{figure*}

\section{The FIRST instrument at the Subaru Telescope}
\label{Sec:first-at-Subaru}

\subsection{Principle of the FIRST instrument}
FIRST aims at using a single telescope as a coherent interferometer. The  instrument concept is shown in Fig.~\ref{fig:first-principle}. Similarly to the aperture masking technique, FIRST samples portions of the pupil and recombines the corresponding beams interferometrically. The use of single-mode fibers, in which the \rev{subpupil} beams are injected, allows for the remapping of the pupil from a redundant array (input pupil) to a \rev{nonredundant} array (output pupil). By nature, single-mode fibers filter out high order perturbations of the wavefront over each \rev{subaperture}, removing speckle noise and providing an increased fringe contrast. The \rev{nonredundant} recombination of $N$~\rev{subapertures} gives access to $N \times (N-1) / 2$~independent measurements\revision{, the complex coherences,} present in the original pupil. Finally, a spectrometer adds spectral information to the recombined interferometric signal. Based on the analysis of the fringes as a function of wavelength, we can derive the optical path difference (OPD) between the different pairs of \rev{subpupils}, as well as CPs.

\subsection{A \rev{submodule} of SCExAO}

The Subaru Coronagraphic Extreme Adaptive Optics~\citep[SCExAO - ][]{2015PASP..127..890J} instrument is located on the Infra-Red (IR) Nasmyth platform of the Subaru Telescope. It is fed by AO188~\citep{minowa2010performance}, an adaptive optics system featuring a 188-element curvature sensor that delivers the first stage of wavefront correction. SCExAO's primary purpose is to perform additional wavefront correction and starlight suppression thanks to coronagraphy, in order to study circumstellar environments~\citep{Currie_2018, vampires, kotani2020reach}.

\begin{figure}[!t]
    \centering
    \includegraphics[width=\linewidth]{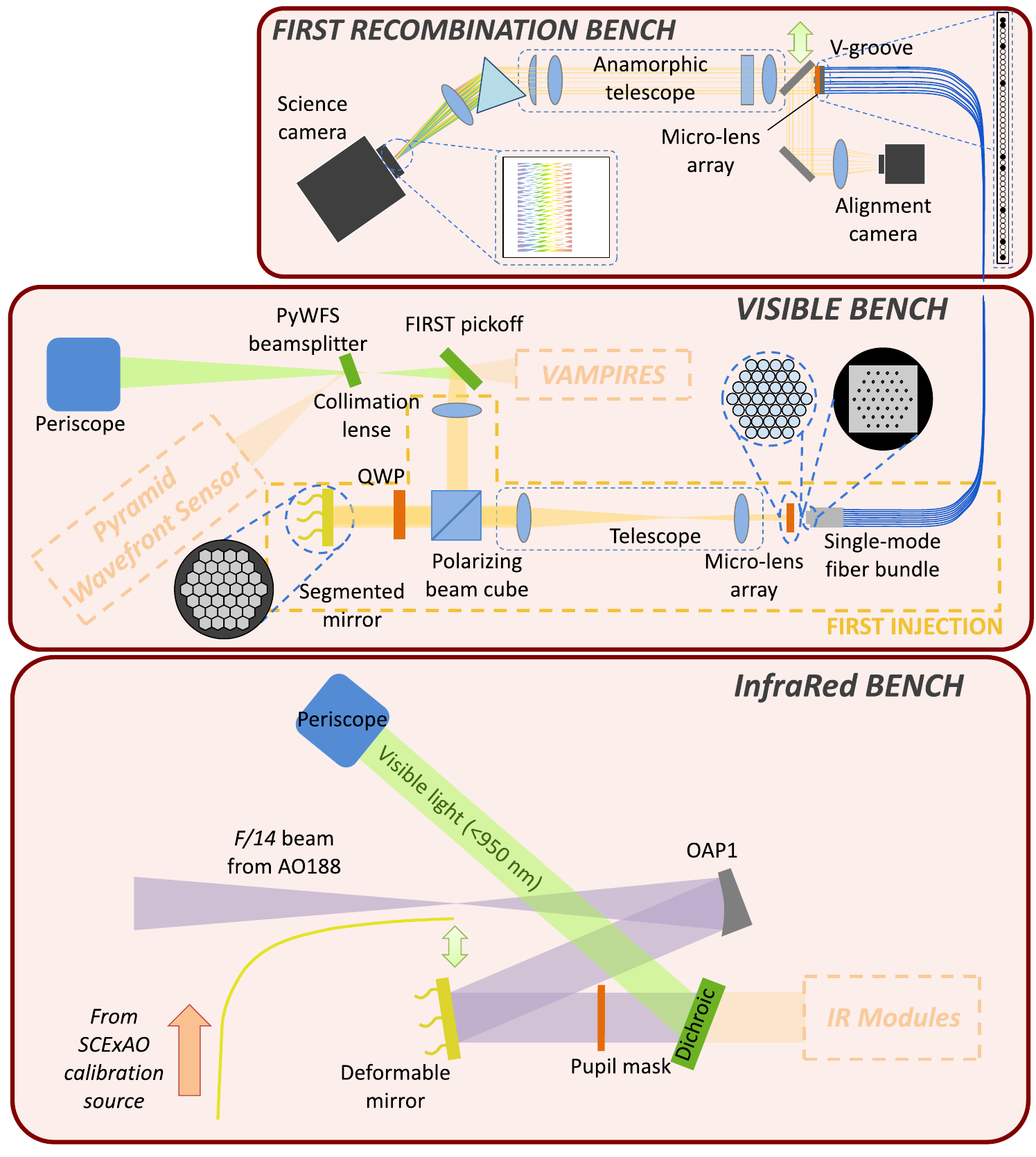}
    \caption{SCExAO optical path from the entrance (bottom) to the FIRST injection module sitting on the visible bench (middle), and in the FIRST recombination bench (top).}
    \label{fig:first-scexao-scheme}
\end{figure}

As can be seen in Fig.~\ref{fig:first-scexao-scheme}, the light reflected off the SCExAO deformable mirror (DM - 2\,040 actuators) is split by a dichroic beamsplitter. Visible light is reflected \rev{toward} a periscope, sending the light \rev{toward} the upper bench dedicated to the visible wavefront sensor and modules. At the output of the periscope, a first beamsplitter reflects part of the light \rev{toward} the pyramid wavefront sensor~\citep[PyWFS - ][]{Lozi_2019}. We commonly use here a short-pass dichroic cutting at $800$~nm, allowing the PyWFS to operate in the $800-950$~nm band to provide a wavefront quality over 80\% Strehl in H-band. The transmitted beam reaches the FIRST pickoff mirror, which is placed on a motorized stage to allow for light to be sent to either VAMPIRES or FIRST.

{}

\subsection{The FIRST setup}
\label{subsec-FIRSTsetup}

\begin{figure}[!h]
    \centering
    \includegraphics[width=0.95\linewidth]{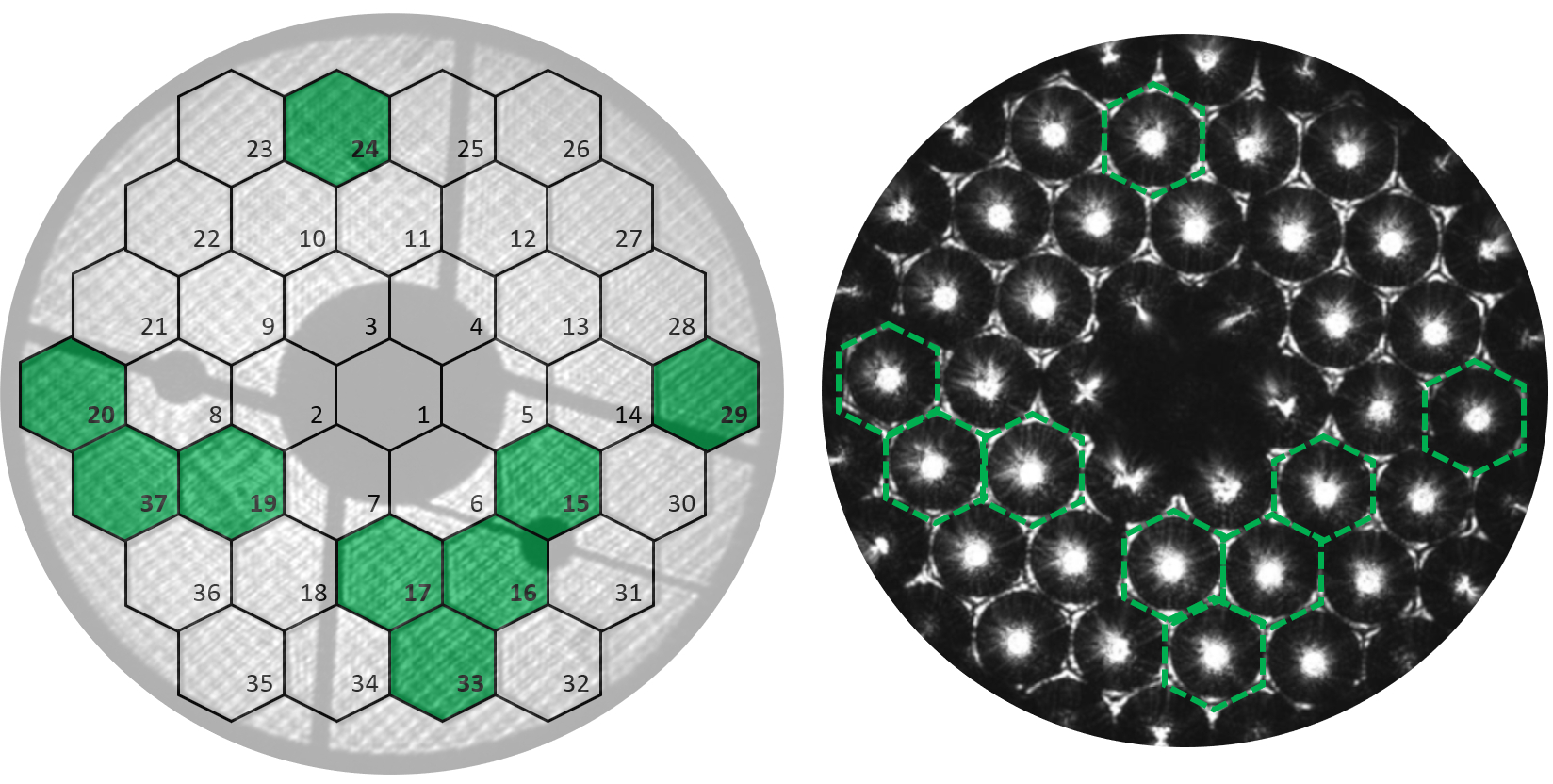}
    \caption{\rev{Configuration of the 9 subapertures in FIRST. Left: Image of FIRST's MEMS conjugated with the Subaru Telescope pupil}. The \rev{subapertures} are numbered after the segment index. Right: Image of the pupil plane after the micro-lens array, where each lens is conjugated to a MEMS segment: the camera detector is conjugated to both the MEMs plane and the microlens array. In green are highlighted the 9 segments used in FIRST.}
    \label{fig:first-pupil-scheme}
\end{figure}

The FIRST optical components span two different benches corresponding to the injection part and the recombination part respectively. The previously mentioned FIRST pickoff mirror sends a f/27.3 diverging beam to the injection setup. A $120$ mm focal length lens collimates the beam, which is then split by a polarizing beamsplitter cube (BSC). The reflected polarization is sent to a Quarter Wave Plate (QWP) right before being reflected at normal incidence off of a 37-element segmented mirror (vendor: Iris AO). This Micro-ElectroMechanical System technology (MEMS) is composed of hexagonal segments, each driven by three actuators allowing Piston, Tip and Tilt control. It is conjugated to the SCExAO pupil plane (see Fig.~\ref{fig:first-pupil-scheme}-left) and used to optimize the injection into the single-mode fibers. The circular polarization changes sign after the reflection on the MEMS and is then transmitted through the QWP and the polarized beamsplitter cube, such that all the flux with the s-polarization is transmitted in the instrument. An afocal system with two lenses (of focal length $85$~mm and $35$~mm, ensuring a magnification factor of $1/2.6$) is used to match the segment size ($606.2~\upmu$m diameter for the circle inscribed in each segment) to the micro-lens diameter ($250~\upmu$m). The micro-lens array (MLA) is conjugated to the MEMS (see Fig.~\ref{fig:first-pupil-scheme}-right). Each of the micro-lens has a $980~\upmu$m focal length and focuses the light from the \rev{subpupils} on the core of single-mode fibers, held in a bundle on a hexagonal pattern grid with a $250~\upmu$m pitch center-to-center. The current FIRST instrument recombines the light from $9$~\rev{subapertures} (see Fig.~\ref{fig:first-pupil-scheme}) each with a $1.04$~m on-sky diameter.  The fibers of the bundle are connected to single-mode fiber extensions \citep[polished by hand for path length matching, ][]{huby2013caracterisation} that run to the recombination bench. 

{}

\begin{figure}[!h]
    \centering
    \includegraphics[width=\linewidth]{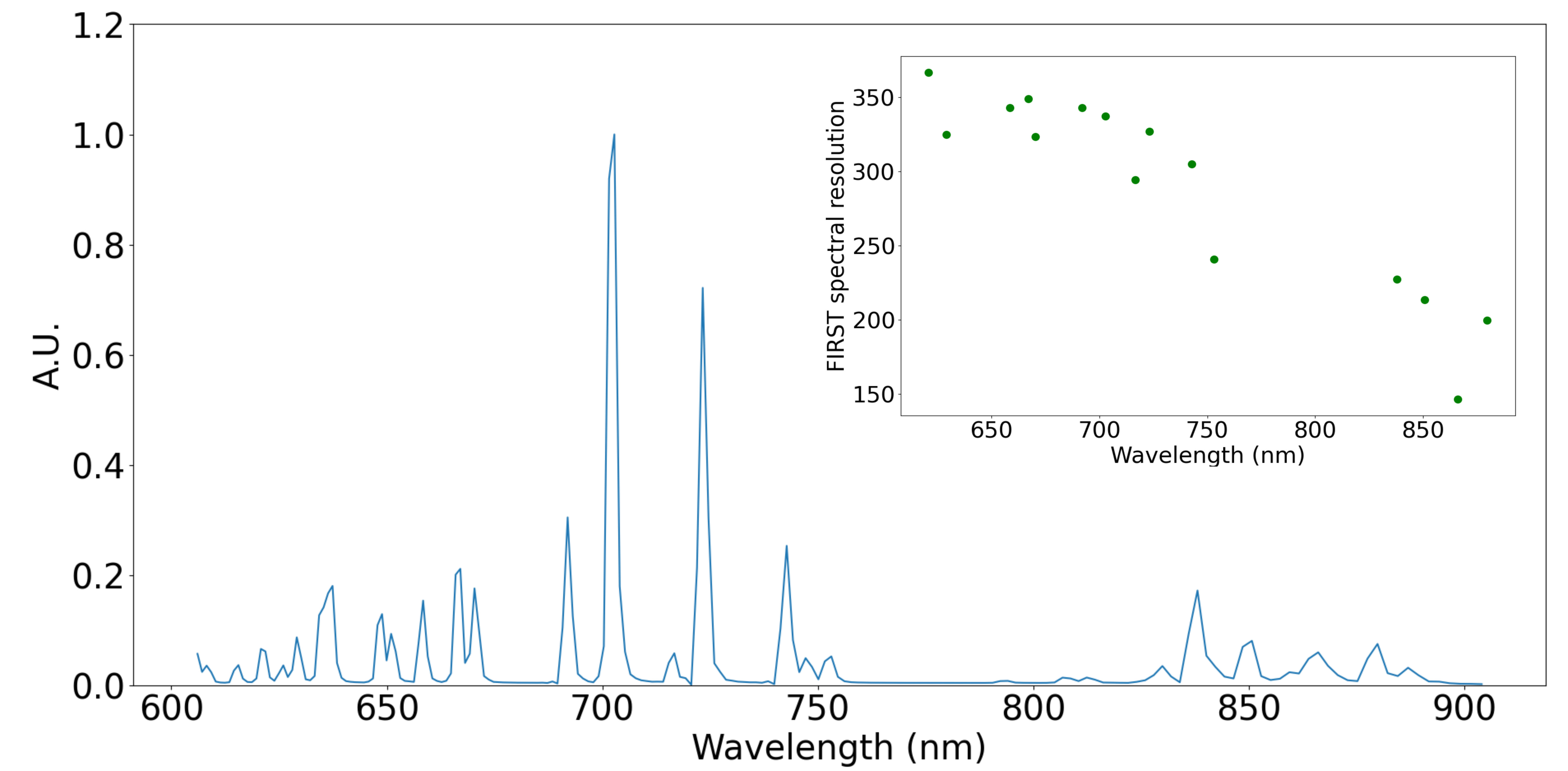}
     \caption{Neon spectrum acquired with FIRST spectrograph. We compute the spectral resolution from the lines acquired in the Neon spectrum and plot its value as a function of the wavelength on the top-right graph.}
    \label{fig:first-spectral-res}
\end{figure}

Inside the recombination bench, the outputs of the fibers are installed on a linear V-groove mount according to a \rev{nonredundant} pattern (pitch of $250~\upmu$m). The beam exiting each fiber is collimated by a micro-lens, and can follow two different paths, that can be selected thanks to a mirror on a motorized stage: (i) \rev{toward} a monitoring camera for the optimization of the coupling efficiency, or (ii) \rev{toward} the spectrometer and the final science camera. The spectrometer consists of an afocal anamorphic system, comprising a set of spherical and cylindrical lenses, that stretches the beam in the dispersion direction and compresses it in the orthogonal direction. The beams are then spectrally dispersed with an equilateral SF2-prism and recombined on the science detector, a $512\times512$ pixel EMCCD (Andor iXon Ultra Life 897). The spectral resolution is estimated using a Neon spectral calibration light source (AvaLight-CAL-Neon-Mini). We present on Fig.~\ref{fig:first-spectral-res} the acquired spectrum. The spectral resolution is calculated by computing the $\lambda/\Delta\lambda$ ratio for each spectral peak (with $\lambda$ and $\Delta\lambda$ the each peak's wavelength and Full Width Half Maximum respectively). The result is presented on the graph in the same figure. We obtain a resolution of  around $300$ at $700$~nm. FIRST's field of view, computed as the size of the diffraction spot from the largest \rev{subaperture}~\citep{guyon2002wide} is about 139~mas at 700~nm.

{}

\subsection{Data \rev{r}eduction pipeline}
\label{Sec:first-datared}
\subsubsection{The variable of interest: \rev{T}he baseline complex coherence}

Classical interferometry data analysis originally relied on studying the fringe pattern Power Spectral Density~\citep{roddier1984long}. This technique is well suited for instruments with large dilution (defined as the ratio between the baseline lengths and \rev{subaperture} radii) since each interferometric term is well isolated. 
The FIRST output pupil being relatively compact, a better suited method is the fringe fitting technique \citep{tatulli2007interferometric}. The data reduction, extensively described in~\cite{huby2012first}, aims at retrieving the complex coherence terms ($\mu_{nn'}$) for each baseline ($n-n'$) from the fringe pattern, written as:
\begin{align}
	\mu_{nn'}=|V_{nn'}|\text{e}^{i\psi_{nn'}}A_nA_{n'}\text{e}^{i\Delta\Phi_{nn'}}\rev{,}\label{eq:mu}
\end{align}
where $|V_{nn'}|$ and $\psi_{nn'}$ are respectively the object's complex visibility modulus and phase. \rev{Furthermore,} $A_n$, $A_{n'}$ and $\Delta\Phi_{nn'}$ are respectively the \rev{subpupil}s $n$ and $n'$ fluxes and differential piston. The differential piston contains the static optical path difference between the fibers added to any aberrations (AO residuals + other hardware related aberrations). The baseline complex coherence terms are derived following the method described in Appendix~\ref{Appendix-datared}. The following \rev{subsection} explains how to retrieve the objects information from the baseline complex coherence terms.

\subsubsection{Closure phase measurements for resolved object characterization}

The main goal of the data reduction pipeline is to provide spatial and spectral information about resolved objects (\rev{e.g.,} stellar binaries or giant stars). This information is contained in the object's complex visibilities, extracted through CP measurements\revision{. It is to be noted that the visibilities cannot be calibrated in our case, and the triple amplitudes are not better suited than CPs for our \rev{t}arget. Therefore, they are discarded. The CPs are, by definition,} the phase of the bispectrum~$\mu_{nn'n''}$:
\begin{equation}\label{bispectrum}
	\mu_{nn'n''}=<\mu_{nn'}\mu_{n'n''}\mu_{nn''}^*>,
\end{equation}
where $n$, $n'$ and $n''$ are the \rev{subpupil} indexes used to form the triangle and $<>$ the average. The nice feature of this quantity is that it allows \rev{differential phase errors to be cancelled} between \rev{subpupils} since they cancel each other out according to Eq.~\ref{eq:mu}:
\begin{equation}\label{bispectrum_angle}
	\text{CP}_{nn'n''}=\text{arg}(\mu_{nn'n''}) = \psi_{nn'}+\psi_{n'n''}-\psi_{nn''}\rev{.}
\end{equation}
The CP measurements are then fitted to a model to estimate the physical parameters of the observed object. In this paper, we only focus on the case of a star and a companion. In this particular case, the object complex visibility \revision{of two unresolved sources} is written as:
\begin{equation}
    V=\frac{1}{1+\rho}(1+\rho\text{e}^{-2i\pi\V{\Delta}\cdot \V{f}}),
    \label{eq:bin-vis}
\end{equation}
with $\rho$ the flux ratio between the two components, $\Delta$ the separation vector ($\alpha$, $\beta$) between the two components, and $\V{f}$ the spatial frequency of the considered baseline. We then perform a fit to the estimated CP with the model. To do so, we want to minimize the $\chi^2$ function defined for each spectral channel as:
\begin{equation}
    \chi^2(\rho,\alpha,\beta) = \sum_{k}^{n_{CP}} \frac{\text{arg}(\mu_{k,estimate}^{} * \mu_{k,model}^*(\rho,\alpha,\beta) )^2}{{\sigma^{k}}^2}\rev{,}
    \label{eq:chi2}
\end{equation}
with $\mu_{k,estimate}$ the $k^{th}$ estimated bispectrum, $\mu_{k,model}$ the $k^{th}$ \rev{modeled} bispectrum, and $\sigma^{k}$ the error on the $k^{th}$ estimated CP. We can then derive the likelihood function from $\chi^2$ as:
\begin{equation}
    \mathcal{L}(\rho,\alpha,\beta) \propto \text{e}^{-\frac{\chi^2(\rho,\alpha,\beta)}{2}}\rev{.}
    \label{eq:likelihood}
\end{equation}
Each parameter can then be retrieved by marginalizing $ \mathcal{L}$ over all the other parameters.

\subsection{Characterization of the instrument}
\label{Sec:first-characterization}

\subsubsection{Fiber injection and field of view}

\begin{figure}[!h]
	\centering
	\includegraphics[width=0.9\linewidth]{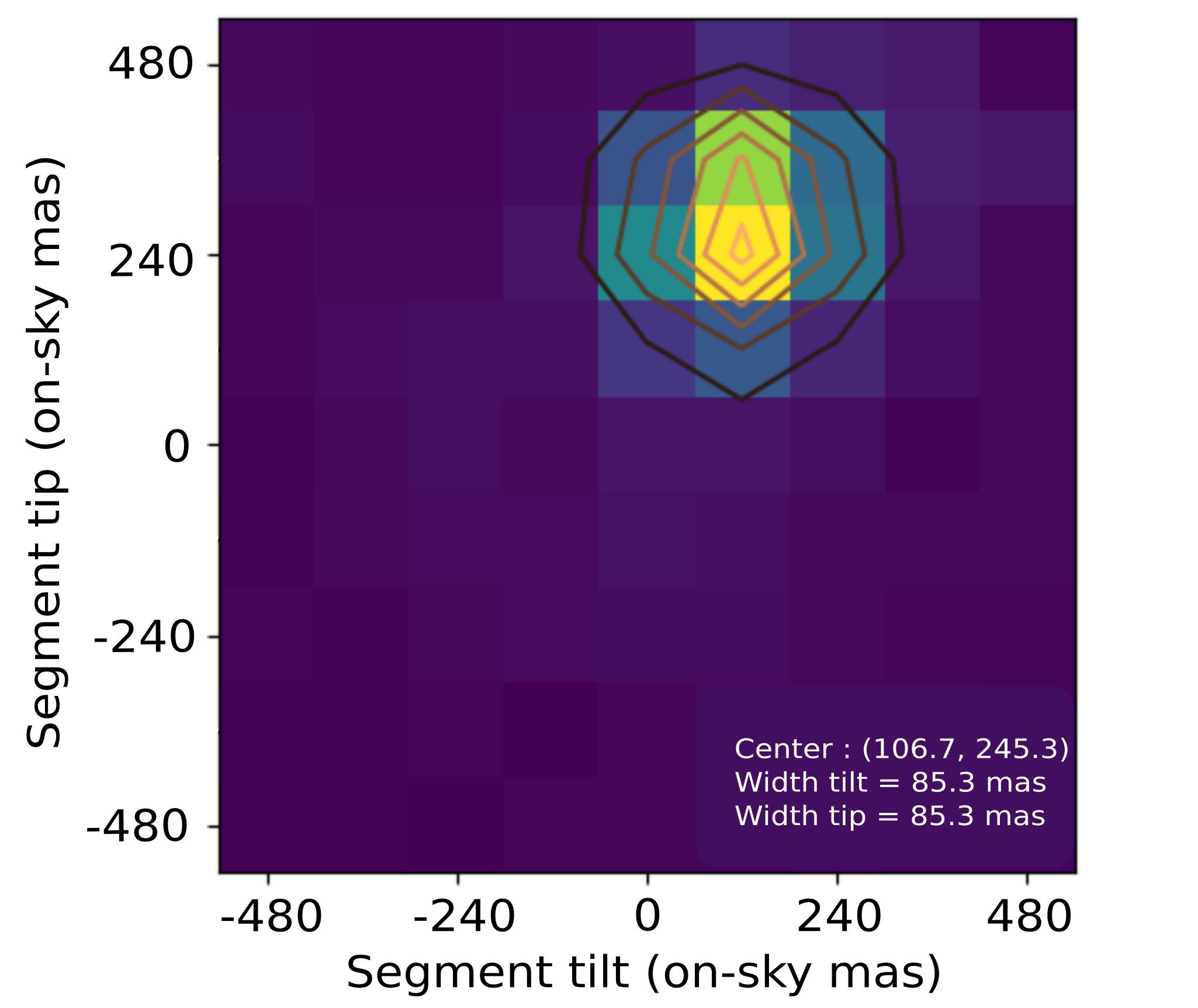}\\
	\caption{On-sky optimization map on \rev{subaperture} associated with the MEMS segment number $16$, with an integration time of $2$~ms for each of the 81 individual images acquired to derive the map. The tip-tilt range is $-2$~mrad to $+2$~mrad on the segment, with a $0.5$~mrad step. This corresponds to $\pm480$~mas on-sky, with a step of $106$~mas.}
	\label{fig-optim-map}
\end{figure}

A key step in the FIRST operation is the optimization of each \rev{subaperture} signal injection into the single-mode fibers. We perform the injection optimization using \rev{tip-tilt} actuation of the MEMS segments, each of them being conjugated with a microlens that focuses the light onto a fiber core (see Sect.~\ref{subsec-FIRSTsetup}). We conduct a \rev{tip-tilt} scan of the \rev{nine} segments and measure the flux transmitted by each fiber using the alignment camera (see Fig.~\ref{fig:first-scexao-scheme}) where the \rev{nine} fiber outputs are simultaneously imaged.
We then build what we name optimization maps (one per segment-fiber association), displaying the integrated flux for each segment position of the scanned grid. An optimization map example is presented in Fig.~\ref{fig-optim-map}, acquired on-sky (during the observation of the unresolved star $\rho$~Persei, on September 16th, 2020) for the MEMS segment 16. The tip (or tilt) applied on each segment ranges typically from $-2$~mrad to $+2$~mrad with a step of $0.5$~mrad. Given the size of each \rev{subaperture} on-sky ($1.04$~m) and the size of each segment ($606.2$~$\upmu$m diameter for the inscribed circle), this corresponds to an on-sky tip (or tilt) range of about $\pm480$ mas with a step of $106$ mas. We fit the optimization map signal with a 2D Gaussian function to estimate the best segment position, with \rev{substep} precision. The optimization map is computed on-sky for all segment-fiber associations, simultaneously. The full procedure takes a couple of minutes. 

{}

\begin{figure}[!h]
	\centering
	\includegraphics[width=\linewidth]{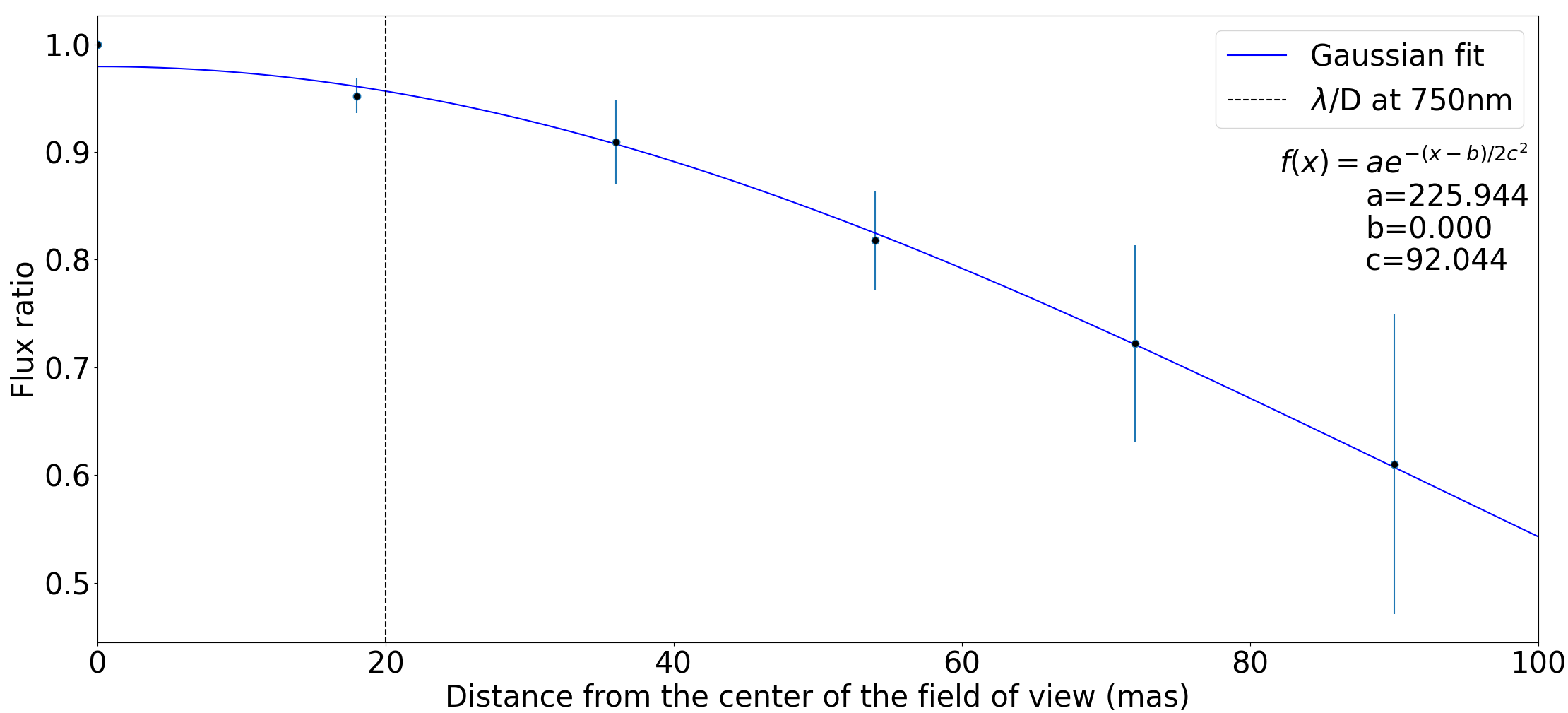}\\
	\caption{Flux losses as a function of the distance from the center of the field of view. Data taken on the SCExAO internal calibration source.}
	\label{fig:fov_flux_loss}
\end{figure}

One of FIRST goals is to detect binary objects. In practice, we assume that there is an object in the center of the field of view and a second off-axis. We want to experimentally assess the flux loss in the case of an off-axis source. To do so, we optimize the fiber injection using the MEMS for a point source simulated by the internal source of SCExAO, then we move the latter in the field of view at various distances. We record the total flux on the photometric camera (from all \rev{subaperture}s) moving the source in 4~directions to $18$~mas, $36$~mas, $54$~mas, $72$~mas and $90$~mas from the center. We then average the 4~direction values to obtain one value per distance. The error on each value is computed as the standard deviation of the 4~values. In Fig.~\ref{fig:fov_flux_loss}, we plot the flux ratio between the averaged off-axis flux and the on-axis flux.

{}

We see that the flux ratio decreases with the distance from the center of the field of view. We display the $\lambda/D$ limit on the graph, since FIRST's main interest lies in its ability to probe regions below the telescope diffraction limit \rev{which is} $20$~mas at $750$~nm. We see that the flux loss for a source located at a distance smaller than $20$~mas is below $5~\%$.

\begin{figure*}[!h]
	\centering
	\includegraphics[width=\linewidth]{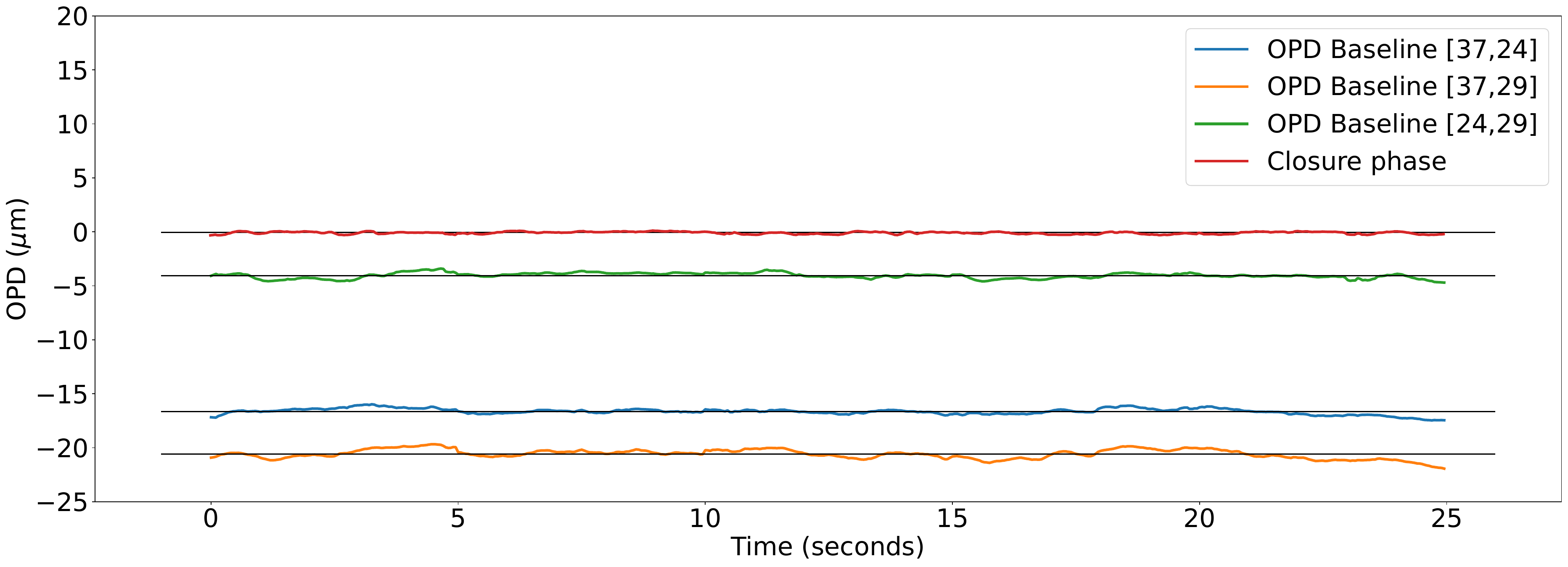}\\
	\caption{Optical \rev{p}ath \rev{d}ifference between 3~\rev{subaperture}s, displayed over the 500~frames of $50$~ms each. We compute the closure phase for consistency check. The bias in the CP is of instrumental origin. Data taken on the SCExAO internal calibration source. }
	\label{OPD_segs}
\end{figure*}

\subsubsection{Throughput and injection efficiency}

In order to assess the throughput efficiency of the instrument, we measured the optical power of the beam using the SCExAO calibration source using a power meter (calibrated for $635$~nm) at different locations in the setup. Table~\ref{table_throughput_2} presents the throughput results.

\begin{table}[!h]
	\centering
\caption{Measured throughput efficiency of the main optical elements in the injection setup.}
\label{table_throughput_2}
\begin{tabular}{ccc}
	\hline
	Optical element & Measured throughput \\ \hline \hline
	Collimation lens & 95\% \\ \hline
    Polarizing BSC + QWP  & \\
    (double path) & 42\% \\ \hline
	Microlens array & 92\%  \\
	\hline        
\end{tabular}
\end{table}

The transmission of $42\%$ obtained after the double passage through the BSC~+~QWP accounts for the selection of one of the two polarizations, since the p-polarization is necessarily dropped. Polarization separation is anyway required at some point in the setup, to avoid the loss of contrast due to the differential optical path difference between the polarizations, induced by birefringence of the polarization maintaining fibers. The $8\%$ difference between the expected $50\%$ for a single polarization and the $42\%$ measured is a loss due to optics surface transmission and quality (cube and QWP). In the near future, we will improve the setup by replacing the BSC+QWP with a D-mirror and quasi-normal incidence on the MEMs, thus reducing the number of optical surfaces. Furthermore, the polarization separation could be performed with a Wollaston prism in the recombination setup.

In order to determine the injection efficiency, the amount of flux incident on each fiber input has been estimated based on an image of the focal plane of the MLA (see Fig.~\ref{fig:first-pupil-scheme}-right). The total flux was measured by the power meter, while the image was used to estimate the fraction of the total flux falling on each individual fiber by integrating the signal in a selected area around the focal point of each microlens. The flux at the output of each fiber of the bundle is then measured with the power meter to evaluate the injection efficiency per segment/fiber. The results are reported in Table~\ref{table_throughput2}. 

\begin{table}[!h]
\centering
	\caption{Estimated injection efficiency in each of the 9 single-mode fibers, obtained on the SCExAO internal calibration source.}
	\label{table_throughput2}
	\begin{tabular}{ccc}
		\hline \hline
		Segment $\#$ & \commentelsa{Flux fraction} & Injection efficiency \\
		\hline
		15           & $1.7\%$                   & $25\%$ \\
		16           & $1.8\%$                   & $41\%$ \\
 		17           & $2.0\%$                   & $69\%$ \\
   		19           & $2.1\%$                   & $56\%$ \\
     	20           & $1.7\%$                   & $46\%$ \\
      	24           & $1.5\%$                   & $37\%$ \\
        29           & $1.1\%$                   & $62\%$ \\
		33           & $1.6\%$                   & $68\%$ \\
		37           & $1.8\%$                   & $52\%$ \\
		\hline \hline      
	\end{tabular}
\end{table}

Fractions of a few percent were expected, since each microlens focuses the light from an area corresponding to less than 1/37 of the whole pupil area, leading to a maximal fraction of 2.6$\%$. Lower fractions were measured, due to possible vignetting by the edges of the pupil or spiders, and to diffraction effects. It is to be noted that segment $\#15$ is \rev{nonfunctioning}, hence the injection could not be optimized\rev{. Therefore, we dismiss this segment in the following and only consider the remaining eight}. The coupling efficiency measured for segment $\#24$ is significantly lower than expected since it is not very close to the edge or near a spider - this could be due to damages on the single-mode fiber itself.

\subsubsection{Phase stability}

We described in Section~\ref{Sec:first-datared} the data reduction pipeline, aiming at estimating each baseline's complex coherence to compute the CP, useful to retrieve spatial information of the observed object. 
In the case of a point source, the objects complex visibility modulus is $1$, and its phase is $0$, and Eq.~\ref{eq:mu} becomes:

\begin{align}
    \mu_{nn'}=A_nA_{n'}\text{e}^{i\Delta\Phi_{nn'}} \text{\hspace{0.5cm} with \hspace{0.5cm}} \Delta\Phi_{nn'} = 2\pi\sigma\delta_{nn'} \text{  } [2\pi]\rev{,}
\end{align}
 
where $\sigma$ is the wavenumber in nm\textsuperscript{-1} and $\delta_{nn'}$ is the OPD in nanometers. We focus here on the characterization of $\delta_{nn'}$. To do so, we acquire data on the SCExAO internal calibration source. We obtain 500~frames of $50$~ms exposure each. Using the data reduction pipeline, we extract the phase from the baseline complex coherences. 
In order to extract the OPD value, we first need to unwrap the phase, fit a linear polynomial, and extract the slope of the latter following:
\begin{equation}
    \frac{\partial\Delta\Phi_{nn'}}{\partial\sigma} = 2\pi\delta_{nn'}\rev{.}
\end{equation}

This operation is done for every baseline, and every frame. Figure~\ref{OPD_segs} shows the evolution of the OPD for three different baselines formed by \rev{subaperture}s $37-24-33$, over the 500 frames. For each baseline, the OPD varies around an average value that is plotted with a solid black line. These variations are caused by turbulence on the SCExAO bench plus thermal/mechanical instabilities. We assume that these variations average to zero over the 500 frames. \rev{This allows us} to state that the OPD average value corresponds to the static path length mismatch between the fibers. The measured standard deviation for the $[37,24]$, $[37,29]$ and $[24,29]$ baselines are respectively $0.3$, $0.4$ and $0.3~\upmu$m. The average OPD variations due to bench turbulence and fiber thermal/mechanical instabilities are contained within $0.4~\upmu$m~RMS. We confirm that the computation of the CP (in red on the graph) cancels out all the perturbations: the average value of the CP is $0.4~\mu$m and the standard deviation is $0.1~\mu$m.

\begin{figure}[!h]
	\centering
	\includegraphics[width=1.\linewidth]{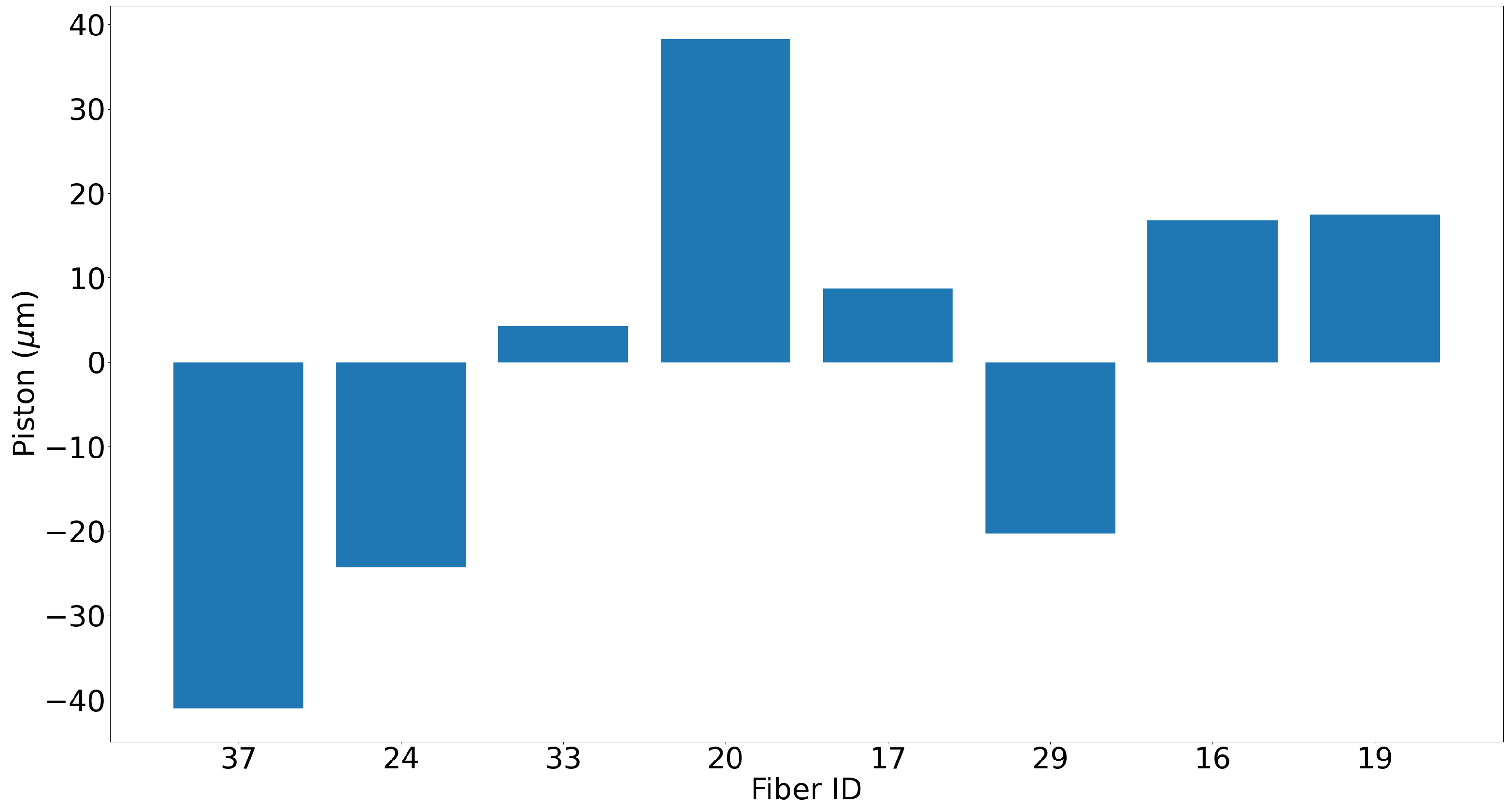}\\
	\caption{Piston value per fiber, corresponding to the path length mismatch between the fibers.}
	\label{P_fib}
\end{figure}

From the OPD measurements, we can compute a piston value per fiber. We present these values in Fig.~\ref{P_fib}. In the future, we will incorporate optical delay lines to reduce these piston values. It is also interesting to highlight that one could extract each \rev{subaperture} piston value and use it to monitor the wavefront. Such interferometric wavefront sensing process could for example be useful to help correct for the island effect~\citep{SPIE_2021-Vievard}. The island effect is due to the pupil fragmentation and induces differential piston values in the pupil fragments, to which classic wavefront sensors like the Shack-Hartman or the PyWFS are not sensitive. 

{}

\section{First On-sky demonstration on the Subaru Telescope}
\label{Sec:first-on-sky}

We present in this section the first on-sky tests performed with FIRST on the Subaru Telescope. Two observation programs were conducted: the first one on a point source target, to test and validate our instrument and data reduction process, the second one on a binary star, to demonstrate FIRST's capability to measure relative positioning of the two binary components. A log of all the data acquired for these studies is presented in Table~\ref{table:Log_obs}.  

\begin{table}[h!]
	\centering
	\caption{Observation log. NA = Not Applicable.}
	\label{table:Log_obs}
	\begin{tabular}{ccccc}
		\hline \hline 
		Target                      &  $N_{img}$     & \begin{tabular}{@{}c@{}} $t_{int}$ \\ $(ms)$  \end{tabular}   & EM gain    &   \begin{tabular}{@{}c@{}} Seeing \\ (at 500~nm) \end{tabular}  \\
		\hline \hline 
		         \multicolumn{5}{c}{\textbf{2020-07-13}}                        \\
		\hline
		Keho`oea                         & 5000           & 100              & 50         &  0.6 arcsec \\
		\begin{tabular}{@{}c@{}} SCExAO \\ calibration source \end{tabular}    & 2000           & 100             &  20        &  NA  \\
		\hline
		         \multicolumn{5}{c}{\textbf{2020-09-16}}                      \\
        \hline
		 $\rho$ Persei                  & 10000           & 40             & 300         &  0.6 arcsec \\
		 Hokulei                     & 9500            & 30             & 300         &  0.6 arcsec \\
		\hline
		         \multicolumn{5}{c}{\textbf{2020-09-17}}                       \\
        \hline
		 Hokulei                     & 9500            & 10              & 300         &  0.6 arcsec \\
		 \begin{tabular}{@{}c@{}} SCExAO \\ calibration source \end{tabular}   & 2000            & 10             &   0         &  NA  \\
		\hline \hline
		\end{tabular}
\end{table}

\subsection{Observation of a point source}
\label{sec:Vega}
\subsubsection{Observation of Keho`oea}
As desccribed in Sect.~\ref{Sec:first-datared}, the \revision{best} end product of the FIRST interferometric data reduction are the CP measurements. For an unresolved point source, the CP signal should be zero, for all triangles. However, this is usually not the case in practice, as instrumental biases remain and need to be calibrated. This is why the observation of calibrator stars before and/or after the target of interest is required. The calibrator data allows \rev{one} to compute CP measurements that are subtracted to the target of interest CP measurements thus removing the instrument bias and producing a CP signal originating solely from the target of interest. 


We acquired data on Keho`oea ($\alpha$~Lyrae or Vega, $m_V=1.25$, $m_R=0.1$), during an Engineering night on the 13th of July 2020 UTC at the Subaru Telescope (Proposal ID: S20A-EN13, PI: Olivier Guyon, Support Astronomer:  Sebastien Vievard, Telescope Operator\rev{:} Andrew Neugarten). The data were acquired between 6:50~a.m. and 7:39~a.m. UTC. The conditions reported by the Canada France Hawaii Telescope Weather Tower~\footnote{\rev{Source\rev{:} http://mkwc.ifa.hawaii.edu/current/seeing/}} showed a seeing measurement of $0.6$~arcsec (measured @500~nm). The SCExAO PyWFS pickoff was set to 850~nm Short Pass Filter, meaning that FIRST received light from about 650~nm to about 800 nm (since the PyWFS pickoff is at an angle, the transmitted wavelength cut-off is lower than 850~nm). Because of limited time to perform our tests, we were not able to acquire data on another unresolved target during that night. In order to calibrate our Keho`oea data, we decided to take, during the slewing of the telescope, data on the SCExAO internal calibration source. Although it does not include bias that could be induced by the telescope and AO188, we can at least calibrate the bias originating from SCExAO. 

\subsubsection{\revision{Data analysis\rev{:} from the fringes to the closure phase}}
\label{subsec-data-analysis}
\begin{figure}[!h]
    \centering
    \includegraphics[width=0.9\linewidth]{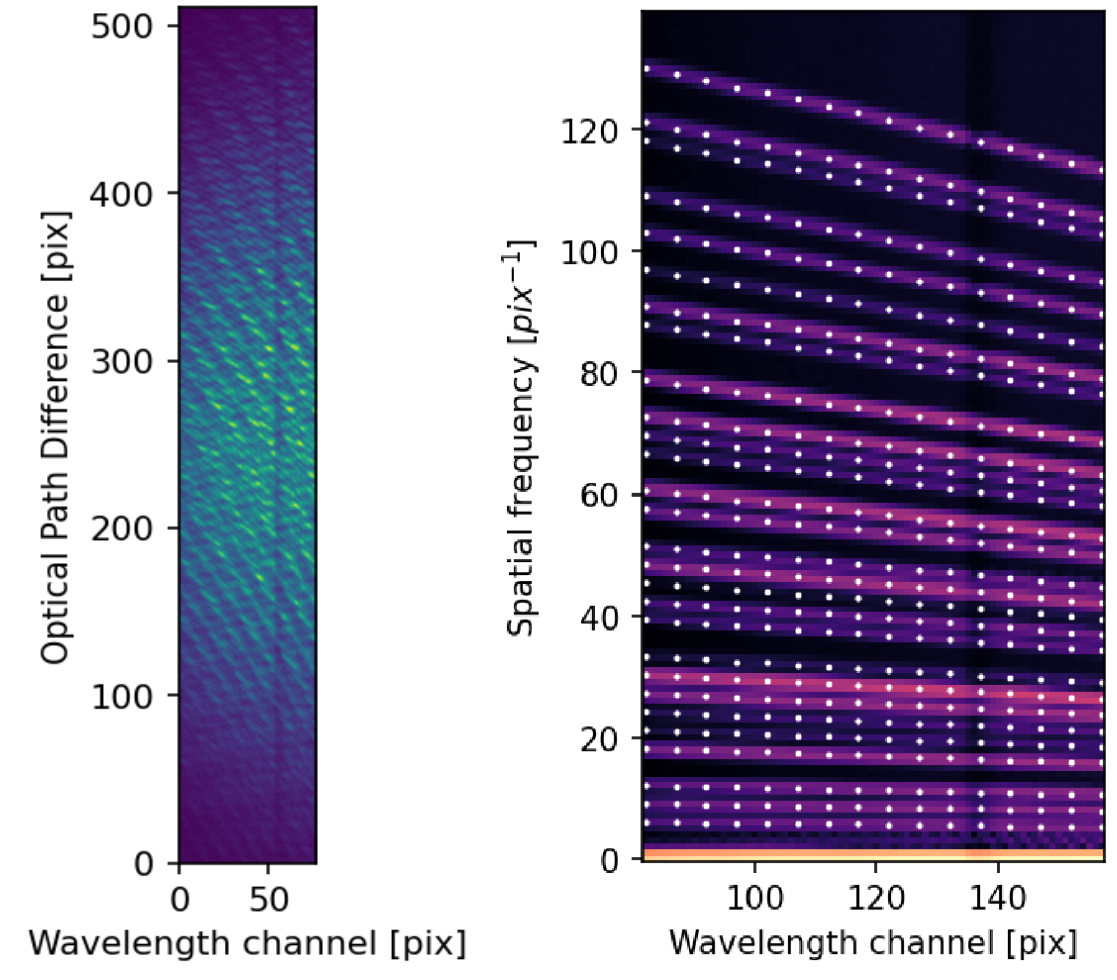}
    \caption{\rev{Images delivered by FIRST.} Left: \rev{Exposure of 100 ms and 50 EMgain} of Keho`oea, corrected from curvature. The vertical dark line around wavelength channel 50 corresponds to the telluric oxygen A band absorption line around 760 to 765 nm. Right: Power spectral density computed from 30 cubes of 100 acquisitions with 100 ms exposure and 50 EMgain. The white dots show the PSD peak fitting for some selected spectral lines.}
    \label{Vega_imDPS}
\end{figure}

Figure~\ref{Vega_imDPS}-left shows a typical Keho`oea image \commentdaniel{of the dispersed interferometric pattern} during this test. The exposure time was 100~ms, and the camera EM gain was set to 50. We acquired a total of 50 cubes of 100 images each. The raw images typically present a curvature caused by astigmatism induced by the dispersing prism~\citep{huby2012firstspie}. This curvature is estimated and corrected thanks to our wavelength calibration data (using the AvaLight-CAL-Neon-Mini Spectral Calibration Light Source). The source is injected in one of the FIRST fibers, and the spectral lines are imaged on the detector. Once the curvature is corrected on each frame, we can compute the Power Spectral Density (PSD). Figure~\ref{Vega_imDPS}-right shows the PSD computed from stacking 30 cubes of the acquired data. From this image, we fit the V-groove step by fitting the peaks for several spectral line. The white dots show the result of the peak fitting. $28$~peaks are present, for each spectral channel fitted, since 8~\rev{subapertures} were actually used during the observations, due to a broken segment on the MEMS.

{}

\begin{figure*}[h!]
    \centering
    \includegraphics[width=\linewidth]{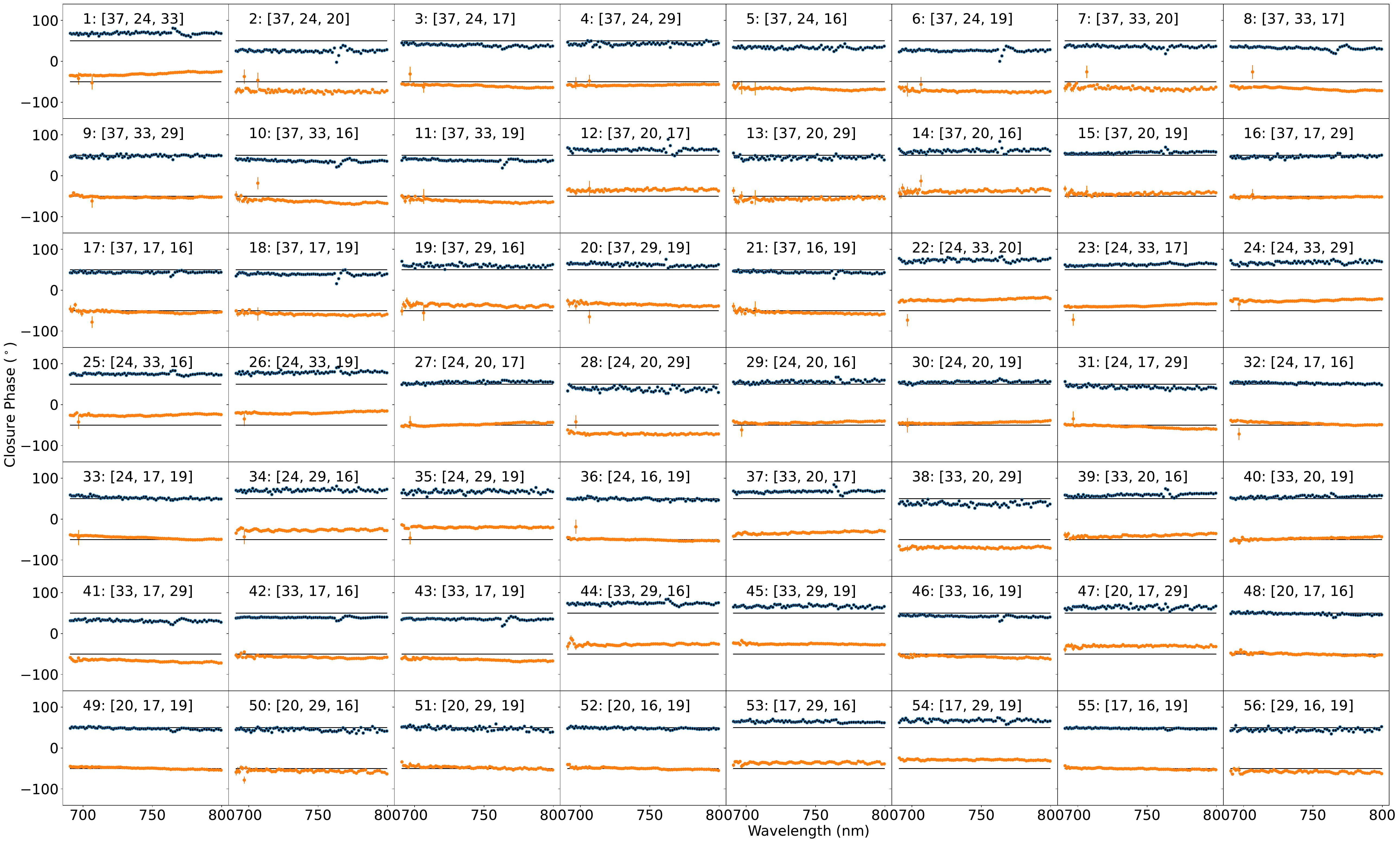}
    \caption{CP measurements for the 56 triangles as a function of the wavelength. The Keho`oea data is displayed in blue, with an offset of $+50^\circ$. The SCExAO internal calibration source data is displayed in orange, with an offset of $-50^\circ$. The triangles are numbered from 1 to 56, with the corresponding \rev{subapertures}.}
    \label{fig:Vega_CPs}
\end{figure*}
\begin{figure*}[h!]
    \centering
    \includegraphics[width=\linewidth]{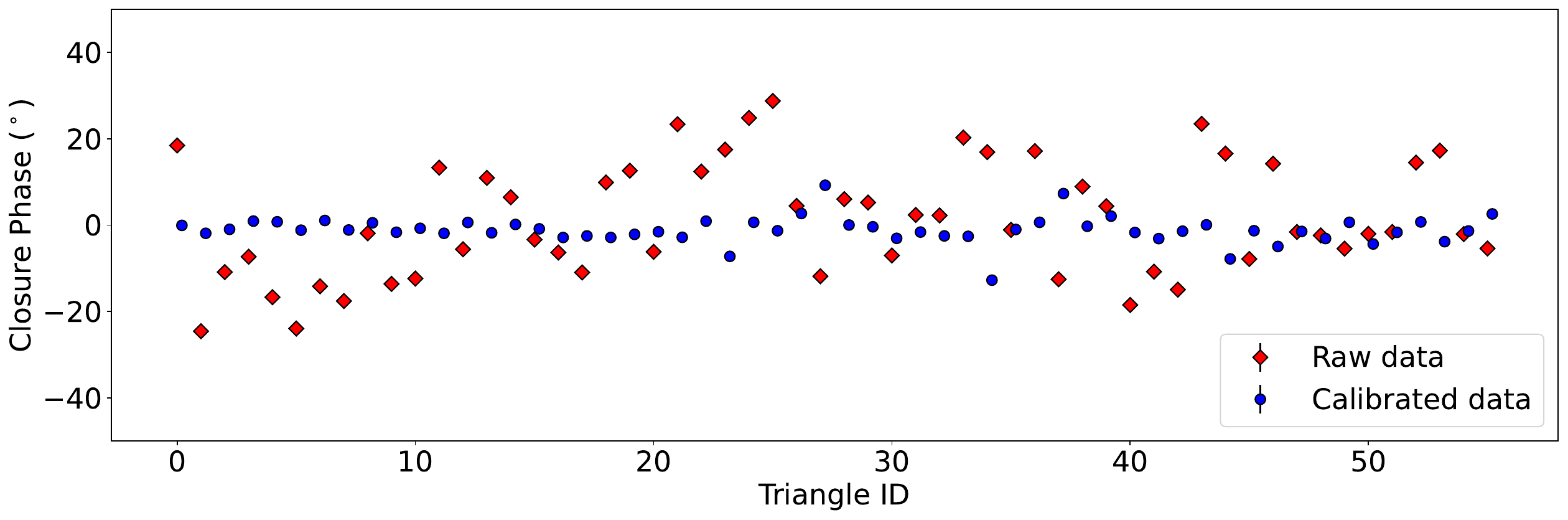}
    \caption{CP computed from the Keho`oea observation, averaged over the wavelength dimension for each triangle. The red diamonds and blue circles respectively correspond to the averaged value before and after calibration using the SCExAO internal calibration source.}
    \label{fig:Vega_CPs_ave}
\end{figure*}

The last required ingredient for the data reduction is referred to as \rev{flat fields}. They correspond to the transmission (flux for each spectral channel) of each fiber alone, or what we denoted as $E_n$ with $n$ the fiber index (see Sect.~\ref{Sec:first-datared}). To acquire these data we move all segments but one in order to "switch off" the fiber signals except for one. This process is repeated eight times to get the eight \rev{flat fields}. We acquire these on each target that we observe. 
We compute our estimates following Eq.~\ref{projection}, to reconstruct the 28 baseline complex coherences from the 8~\rev{subaperture}s.  Finally, following Eqs.~\ref{bispectrum} and~\ref{bispectrum_angle}, we build our 56~CPs. We average the bispectra extracted from the $50\times 100$ images. We also compute 56~CP from the data acquired on the SCExAO internal calibration source (20 cubes of 100 images each, with an exposure time of $100$~ms and an EM gain of 20). Because these two sources are unresolved, their CP values should be $0^\circ$. For \rev{visualization} purposes, we plot the CP measurements from each target on the same graph, but each with a different offset on Fig.~\ref{fig:Vega_CPs}. The CP measurements obtained on Keho`oea are displayed in blue, with an offset of $+50^\circ$ (materialized by a solid black line). The CP measurements obtained on the SCExAO internal source are displayed in orange, with an offset of $-50^\circ$ (materialized by a solid black line). Each of the 56 graphs shows the CP values as a function of the wavelength. 

As expected, some CP values are \rev{nonzero}, due to the instrumental bias. Indeed, the series of blue and orange dots (the CP measurements) do not overlap exactly on the solid black lines (materializing the introduced offset) in all of the 56 graphs. For example, on graph number~1, both CP values are a few degrees above 0. However, for the graph number~49, the bias is very small. We can qualitatively confirm that the bias amplitude for both targets is the same in a first approximation. We note on the Keho`oea data that for some triangles, there is a small "bump" centered around $765$~nm. This is probably due to a decrease of the fringe contrast at this wavelength, induced by the telluric oxygen A band absorption line around 760 to 765~nm. We do not see this bump in the SCExAO calibration source data. Finally, the calibration source data show a disrupted signal on several triangles (2, 7, 12, 13, 14,15) appearing as high frequency (in the wavelength dimension) oscillations. It is a periodic signal, as the function of the wavelength, that we have noticed several times on the SCExAO internal source, but which origin is still uncertain (and under investigations). 

\begin{figure}[!b]
    \centering
    \includegraphics[width=\linewidth]{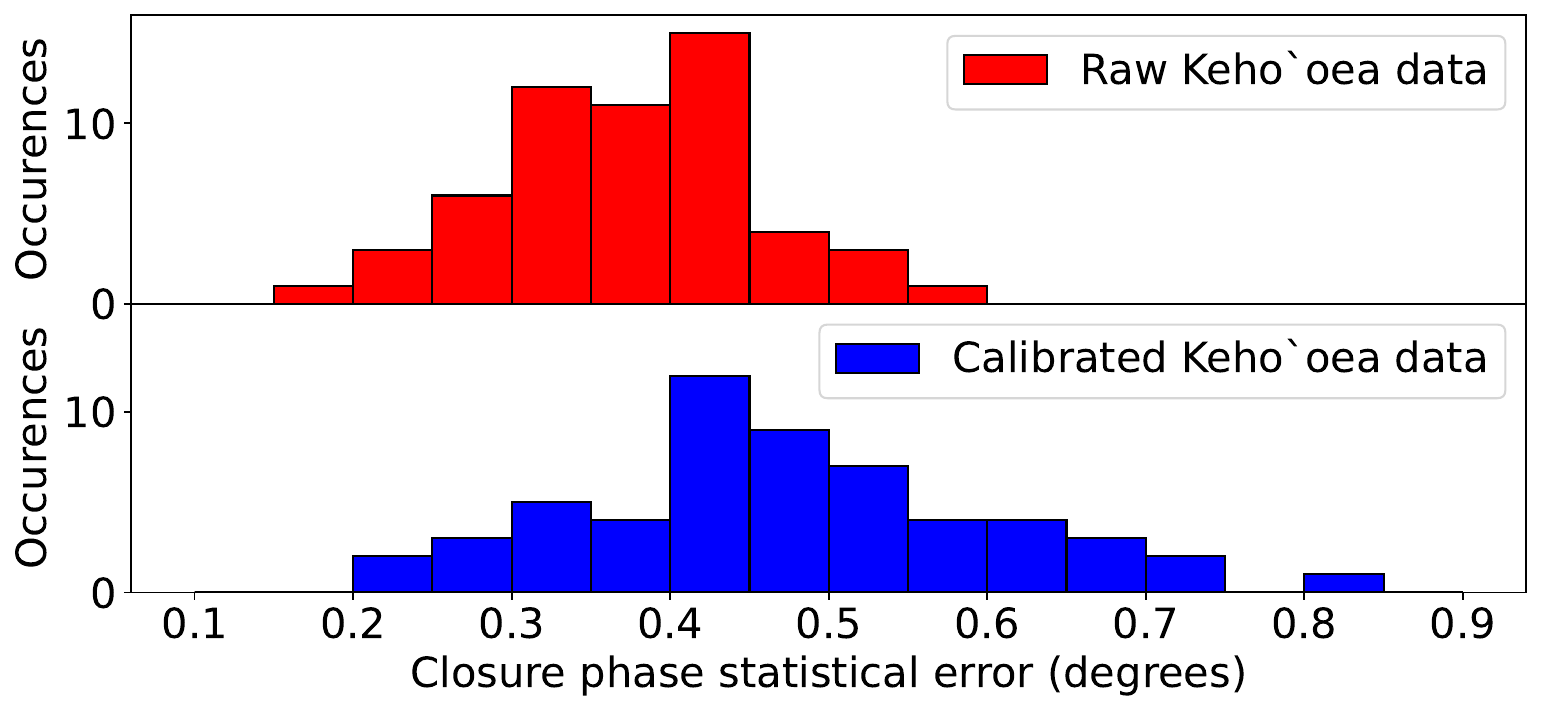}
    \caption{\revision{Histograms of the closure phase error bar estimates, averaged over the spectral dimension. Top and bottom graphs show the histograms of the averaged error bars for the raw CP data and calibrated CP data respectively.}}
    \label{Vega_CP_std}
\end{figure}{}

 \begin{figure*}[h!]
    \centering
    \begin{tabular}{cc}
    \includegraphics[width=0.46\linewidth]{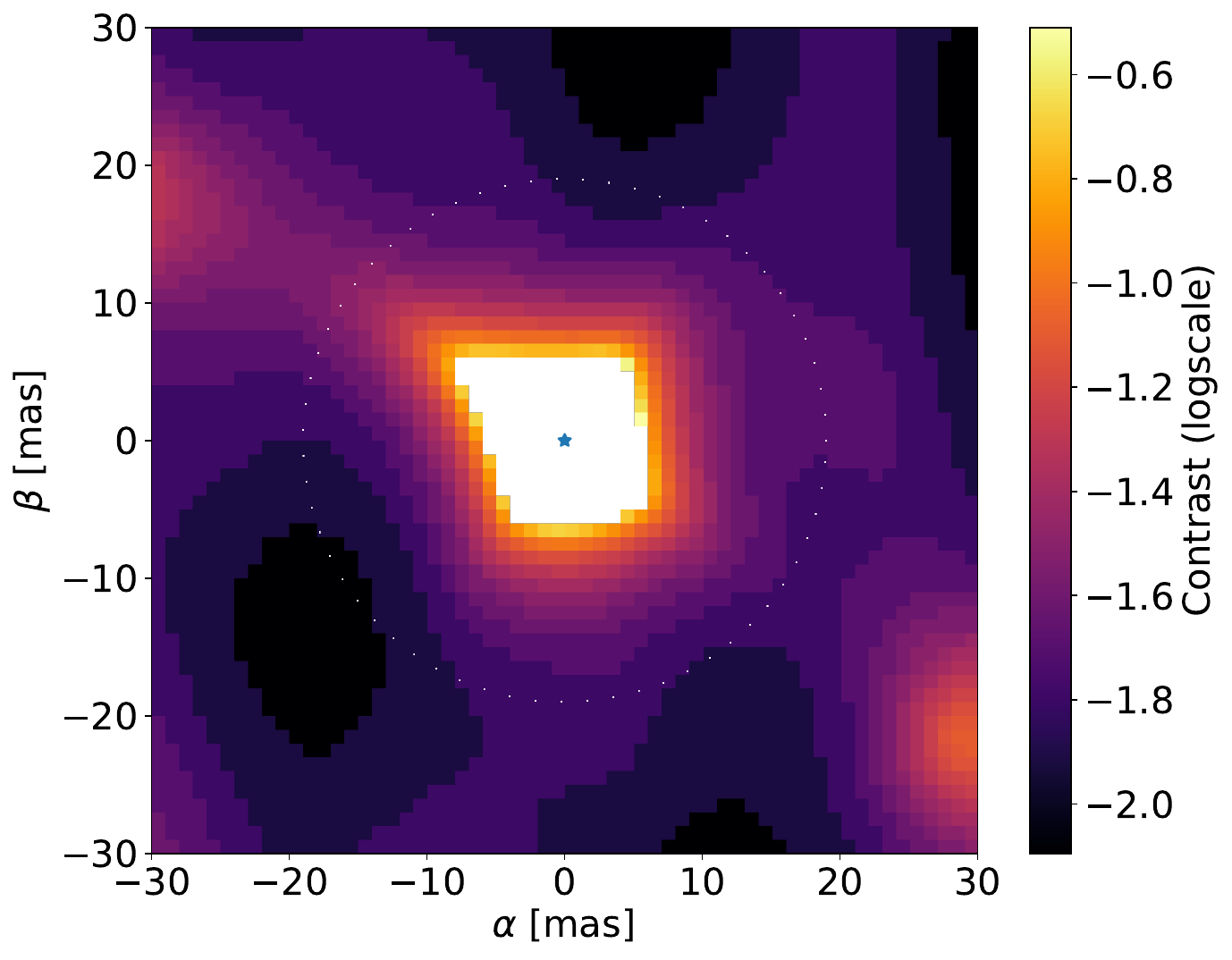} & \includegraphics[width=0.46\linewidth]{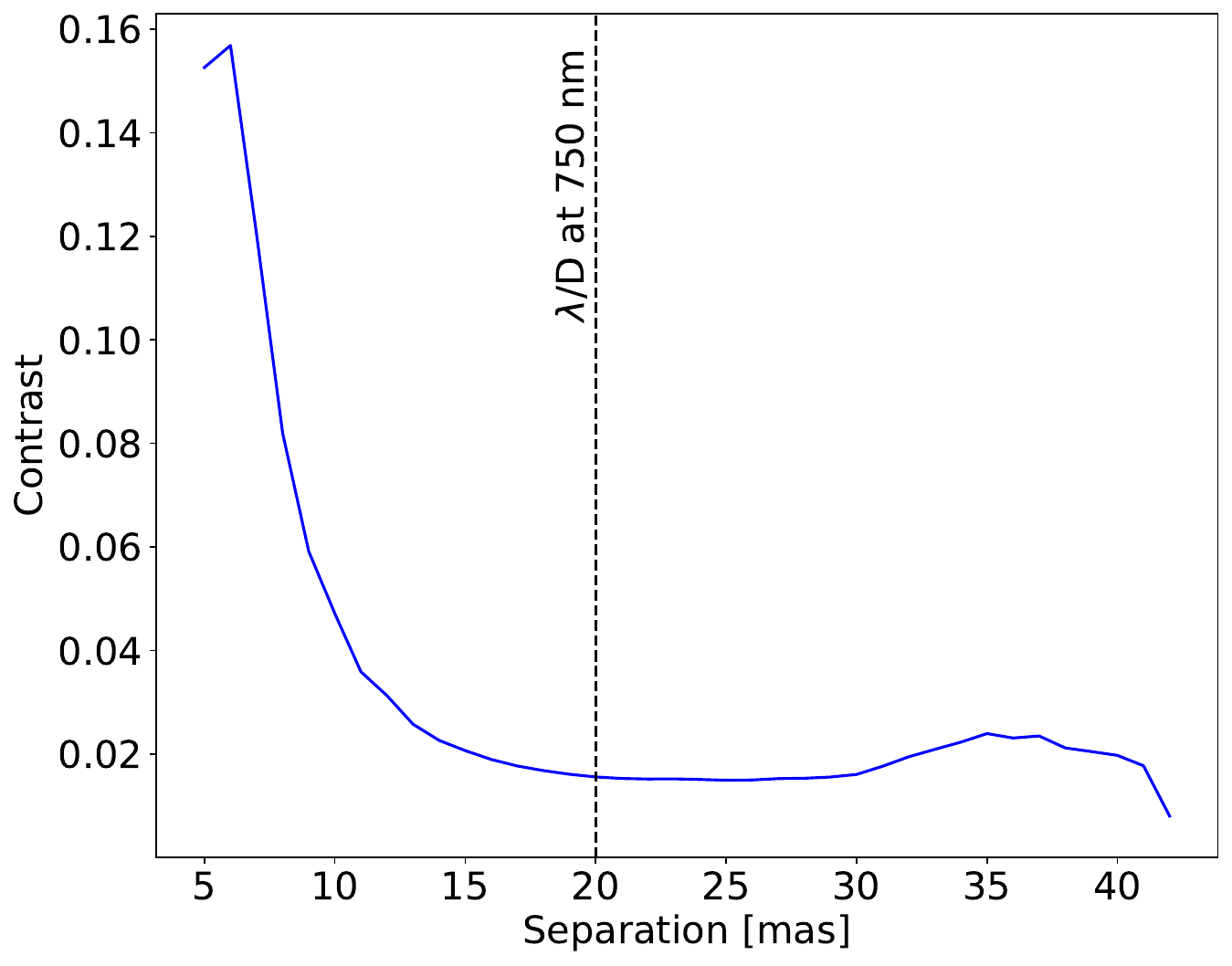} \\
    \end{tabular}
    \caption{\revision{\rev{Left}: Map of $3$-$\sigma$ sensitivity around Keho`oea for the detection of companions, with a total integration time of 500\,s (50 cubes) on target (not including the integration time of a calibrator). \rev{Right}: Computation of the left map radial value (excluding 0-values).}}
    \label{fig:sensitivity}
\end{figure*}

To push the analysis further we compute the average, over the spectral dimension, of the Keho`oea CP signal before and after calibration with the SCExAO internal source. \revision{We also compute the associated statistical error, as the standard deviation over the spectral dimension divided by the square root of the number of spectral channels minus 1.} Results are presented in Fig.~\ref{fig:Vega_CPs_ave}: the graph shows the averaged CP value for each triangle following the numbering of Fig.~\ref{fig:Vega_CPs}. \revision{The error bars illustrating the statistical errors are too small to be seen, so we present them in the form of histograms on Fig.~\ref{Vega_CP_std}.} Before calibration, the average CP spans about $\pm 30^\circ$. After calibration, the average CP values are confined within the $\pm 5^\circ$ range. The averaged value of all CP measurements after calibration is $-0.6^\circ$, with a standard deviation of the averaged values of $0.3^\circ$. \revision{Four} reasons can lead to a \rev{nonzero} value here: \rev{i)} on-sky CPs are calibrated using SCExAO's internal source which does not take into account biases that could originate from the telescope, AO188 and/or turbulence residuals; \rev{ii)} instrumental systematic errors (which could be related to the observed CP oscillations); \rev{iii)} Scattered light from the Keho`oea surrounding debris disk can also induce a signal in the CP; \rev{iv)} Our error bars might be underestimated. Nevertheless, we confirm thanks to this example that our data reduction pipeline is valid.

\revision{Finally, we wish to compare the instrument performance between this observation and the one performed at the Lick Observatory on the same target~\citep{huby2012first}, where uncalibrated and calibrated CP data showed a statistical error of $0.3^\circ$ and $0.5^\circ$ respectively, at best. At the time, closure phase were averaged over 133 spectral channels and the effective integration time was 1000\,s (prior to selection of the best frames). Only 5~closure phase over 56 were exploitable. Figure~\ref{Vega_CP_std} shows the histograms of the statistical errors obtained for the present observation. Uncalibrated and calibrated CP measurements have statistical errors, at best, of about $0.15^\circ$ and $0.2^\circ$ respectively. This corresponds to an improvement in performance by a factor of about 2. FIRST at Subaru also offers several notable enhancements\rev{:} \rev{i)} The acquisition procedure has undergone improvements, making it significantly faster. In our current implementation, the process took less than 10 minutes for an effective integration time of 500 seconds. This represents an upgrade from the previous setup, which required approximately 1.5 hours for an effective integration time of 1000 seconds. \rev{ii)} Unlike before, there was no need here for frame selection thus optimizing the efficiency of the instrument. \rev{iii)} Furthermore, all 56 closure phase measurements were fully exploitable here. This achievement ensures that every single measurement is valuable and contributes to the overall analysis, significantly enhancing the quality of the results.} 

\subsubsection{\revision{Detection limits}}
\label{subsec-det-lim}
\revision{We now use the data previously analyzed to assess FIRST dynamic range detection limit. It consists in estimating, for given positions around Keho`oea, the lowest flux ratio at which a potential companion could have been detected. We \rev{used} the method derived 
 in~\cite{absil2011searching} and \cite{le2012sensitivity}, making use of CP computed on a point source to generate sensitivity maps. We fit the Keho`oea CP measurements to binary models, as described in Section~\ref{Sec:first-datared}, and derive the $\chi^2$ associated for each set of parameters $(\rho,\alpha,\beta)$. We then normalize the $\chi^2(\rho,\alpha,\beta)$ values by dividing them by the $\chi^2$ for $\rho = 0$. This \rev{allowed us} to convert the normalized $\chi^2$ into a probability following: 
 \begin{equation}
     P(\rho,\alpha,\beta) = 1 - CDF_\nu\GP{\frac{\nu\chi^2(\rho,\alpha,\beta)}{\chi^2(\rho=0)}},
 \end{equation}
 where $CDF_\nu$ is the $\chi^2$ cumulative probability distribution function with $\nu$ degrees of freedom. In our case, the number of degrees of freedom is given by the number of independent CP triangles ($21$) multiplied by the number of spectral channels ($77$) minus 1 (total\rev{:} $1616$ degrees of freedom). The tested data set can allow the model with no companion ($\rho=0$) to be rejected if $P$ is below a certain threshold. If this threshold is set at a probability of $0.27\%$, then this ensures a $3$-$\sigma$ detection according to \cite{le2012sensitivity}. For each companion position probed $(\alpha,\beta)$, the sensitivity limit is derived by the lower flux ratio $\rho$ that is not compatible with the single star model ($\rho=0$), \rev{meaning that} P<$0.27\%$. The resulting sensitivity map is shown on Fig.~\ref{fig:sensitivity}-left for the $\pm30$~mas window around the central star. Companions as faint as $8\times 10^{-3}$ times the central star flux can be detected in some area of the field of view. The central part of the map, in white, contains 0-values - meaning that no companion can be detected. From this map, we compute the radial average, leading to Fig.~\ref{fig:sensitivity}-right, where the average sensitivity curve as a function of the separation to the central star is plotted. This performance study shows that a companion could be detected at a the separation of 5\,mas, or $\lambda/4D$. The achievable contrast decreases rapidly from around 0.16 to 0.02 between 5~mas and 15~mas. As expected with interferometric techniques, the contrast limit then reaches a plateau around 0.02 between 15~mas and 40~mas.}
 
 \subsubsection{\revision{Magnitude limit}}

\begin{figure}[!h]
    \centering
    \includegraphics[width=\linewidth]{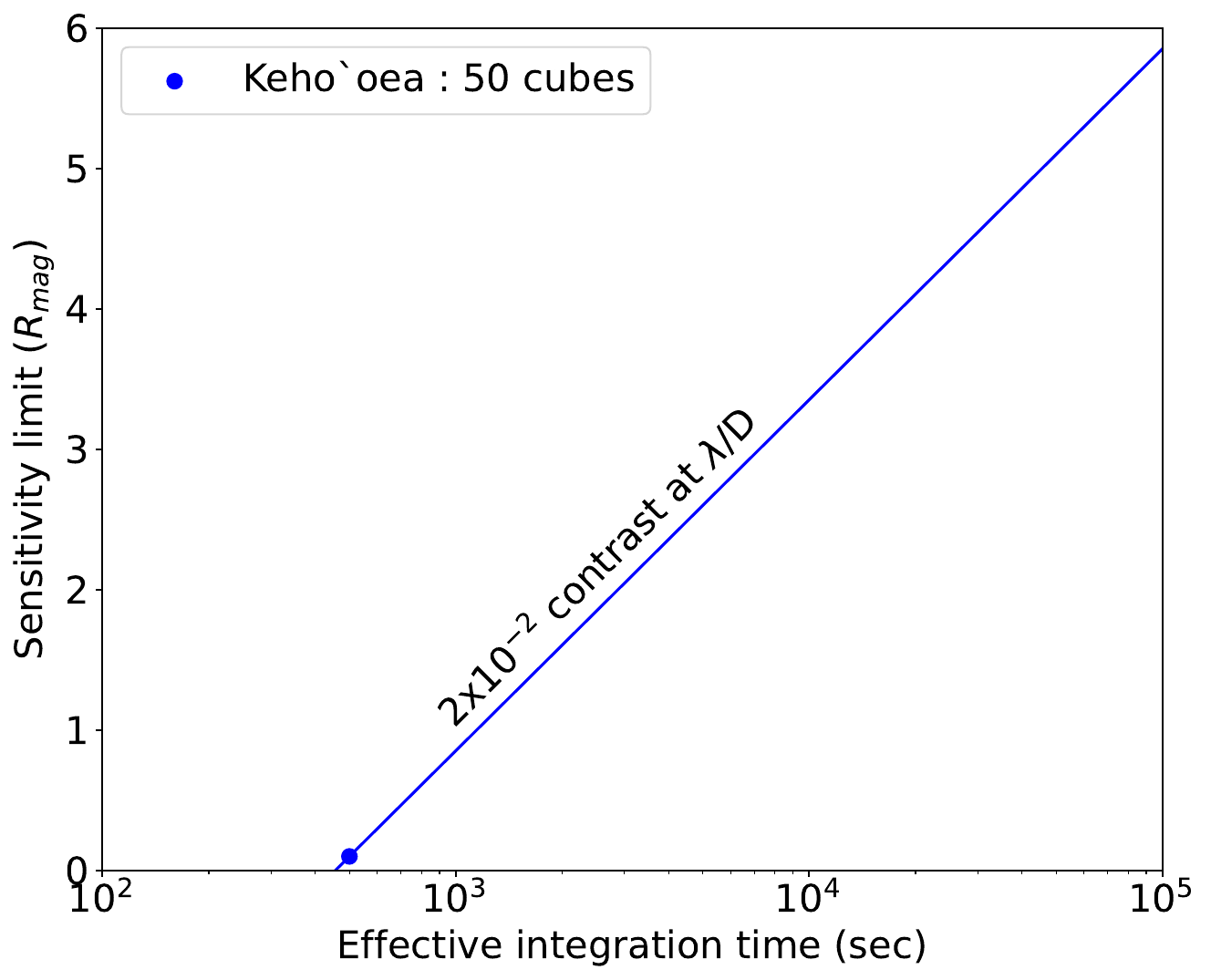}
    \caption{\revision{Extrapolation of the detection performance for different star magnitude as a function of the effective integration time.}}
    \label{fig-maglim}
\end{figure}

\revision{We extrapolate the previous detection limit results obtained on Keho`oea to estimate the instrument performance for stars with different magnitudes. Similarly to~\cite{huby2012first}, we can assess the effective integration time $\tau$, that would be required to obtain the same performance as on Keho`oea ($R_{mag,0} = 0.1$ observed with an integration time $\tau_0$) using\rev{:}
\begin{equation}\label{eq-Rmaglim}
    R_{mag} = R_{mag, 0} + 2.5\text{log}\GP{\frac{\tau}{\tau_0}},
\end{equation}
This extrapolation is valid under the assumptions of 1- similar observation conditions, and 2- photon noise limited regime. 
As presented in the previous section, the contrast limit for a total effective integration time of $\tau_0=500~s$ (50 cubes with 100 images of $100~ms$ each) reaches 0.02 for separations greater than $\lambda/D$. 
This limit corresponding to $R_{mag,0} = 0.1$ is plotted in Figure~\ref{fig-maglim}, and the sensitivity limit is extrapolated to give $R_{mag}$ as a function of the effective integration time $\tau$ required to reach the same contrast performance. }
{}

This study allows \rev{us} to convey that FIRST can, at this stage, reach contrasts of about 0.02, around magnitude 3 stars, with an effective exposure time of about 2~hours.

\subsection{Observation of a binary}
\subsubsection{Observation of Hokulei}

After validation of the data reduction pipeline on a point source, we were able to study the case of a binary system: Hokulei ($\alpha$~Aurigae or Capella, $m_V=0.08$,$m_R=-0.52$). The goal was to measure the separation and position angle (PA) of \correction{the two binary components}. Hokulei is a binary that is well resolved by the Subaru Telescope in the visible, with a $56.4$~mas semi-major axis and a period of about $104$~days~\citep{torres2015capella}. Although this target is not the most scientifically significant for FIRST (since we ultimately aim at targets below the diffraction limit of the telescope), it is an easy engineering target (bright magnitude in the visible and low contrast) to test and validate the CP fitting technique described in Sect.~\ref{Sec:first-datared}.\\

We observed Hokulei during the engineering observing nights on the 16th and 17th of September 2020 UTC (Proposal ID: S20B-EN13, PI: Olivier Guyon, Support Astronomer: Sebastien Vievard, Telescope Operator: Andrew Neugarten) as summarized in Table~\ref{table:Log_obs}. We acquired data with FIRST between 2:43~p.m. and 3:00~p.m. UTC on Sept. 16th 2020, while seeing as reported by the CFHTWT was about $0.6~$arcsec. On Sept. 17th, we acquired data between 3:07~p.m. and 3:36~p.m. UTC, with an average seeing of $0.6$~arcsec as well. The SCExAO PyWFS pickoff was set to 850~nm Short Pass. During the first night, we used $\rho$~Persei ($m_V=3.39$, $m_R=1.59$) as a calibrator. We did not have enough on-sky time during the second night to acquire data on a calibrator. Therefore, we used SCExAO internal calibration source right after observations to calibrate the CP bias. 
The Hokulei expected separation and PA were computed thanks to the Binary Analysis Tools (BATs~\footnote{https://github.com/scexao-org/Binary\_analysis}). BATs uses Hokulei orbit parameters from~\cite{torres2015capella} to estimate the companion position at a given date (see Fig.~\ref{fig:Capella_pos}). 

\begin{figure*}[!h]
    \centering
    \includegraphics[width=0.99\linewidth]{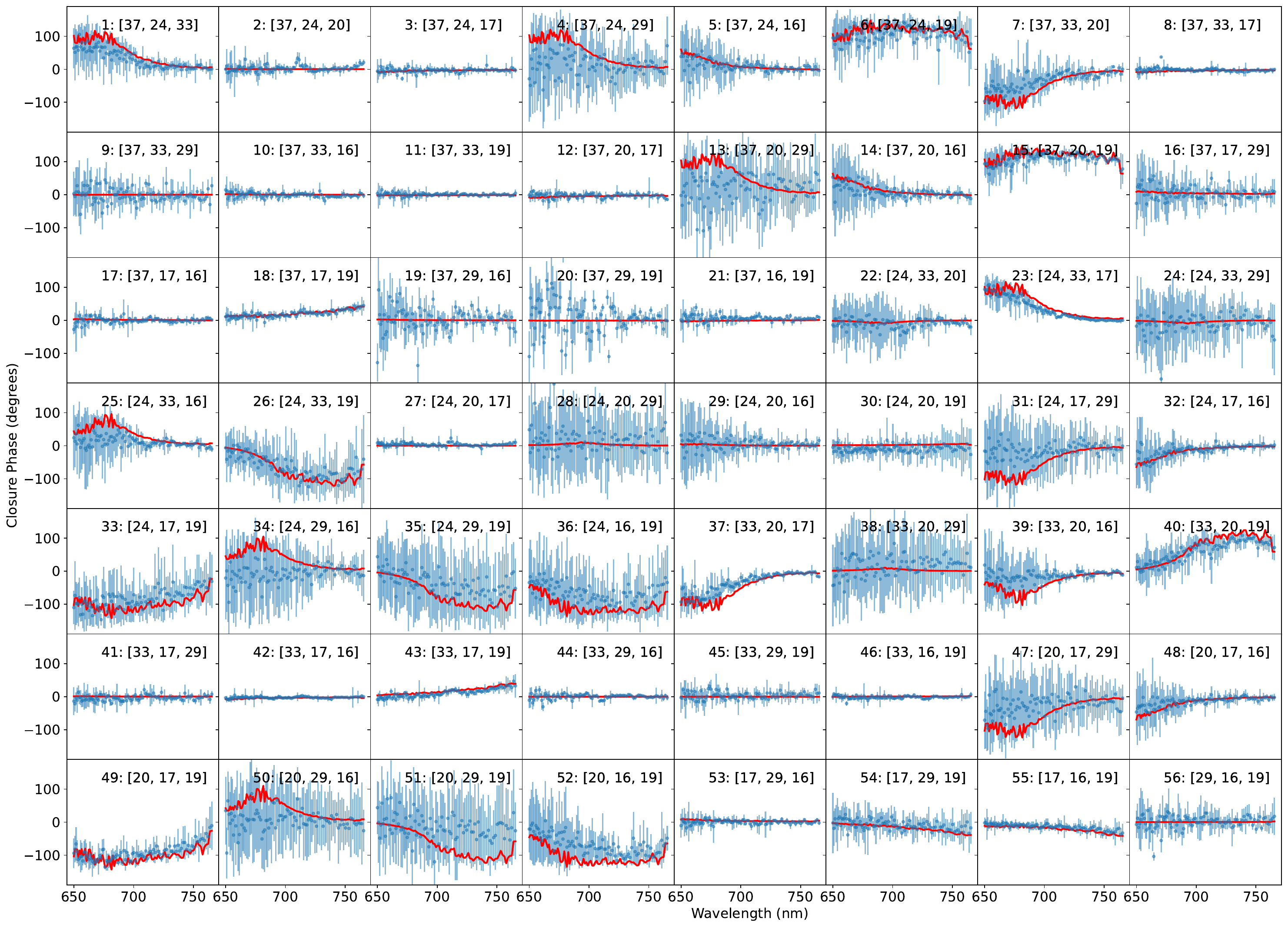}
    \caption{CP fit after calibration, for one cube of data obtained on Hokulei. The blue curves are the 56 CPs measured by FIRST, the red curves are the fitting results.}
    \label{fig:CP_Capella}
\end{figure*}{}

\begin{figure*}[h!]
    \centering
    \begin{tabular}{ccc}
    \includegraphics[width=0.32\linewidth]{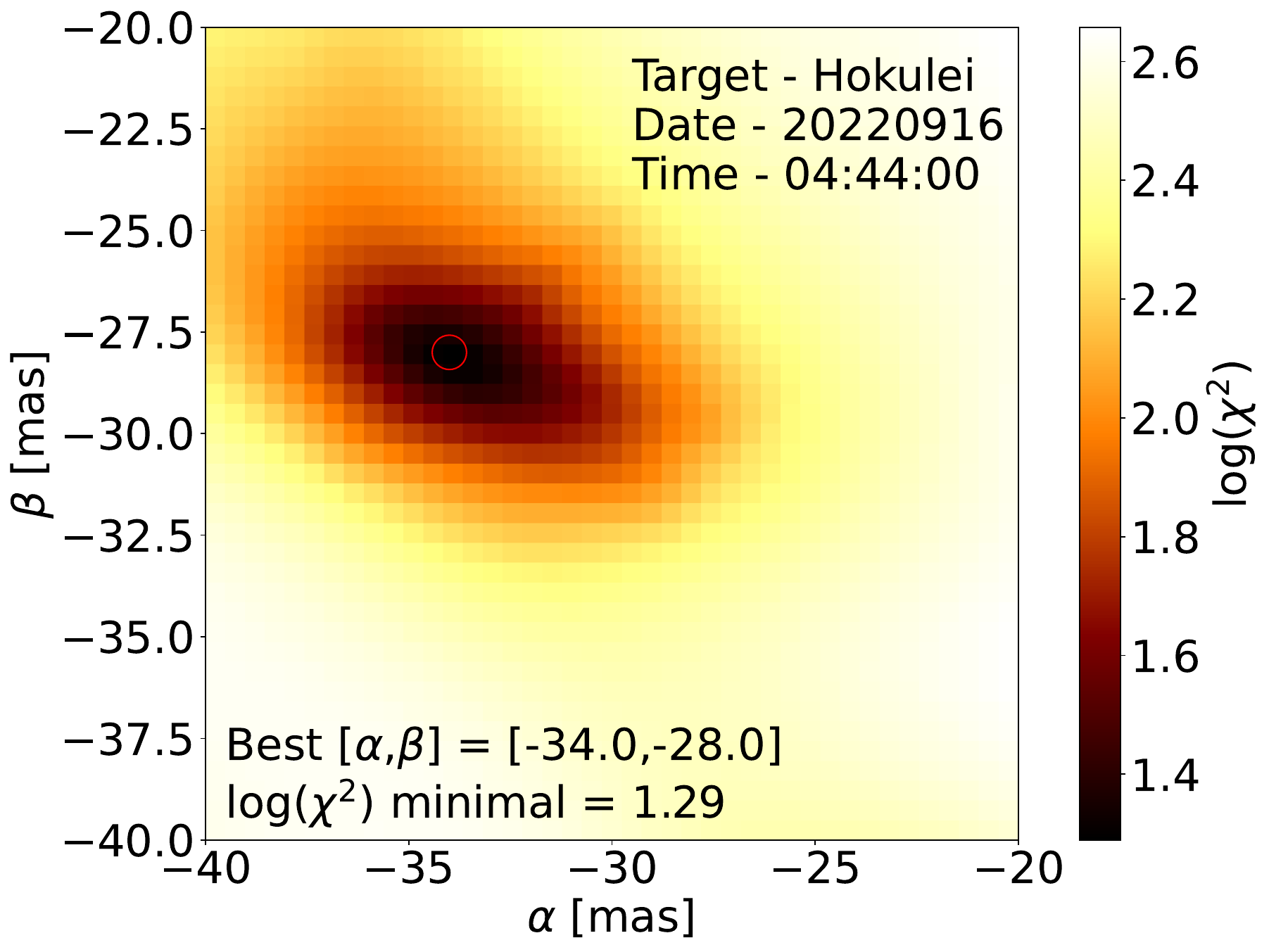} & \includegraphics[width=0.32\linewidth]{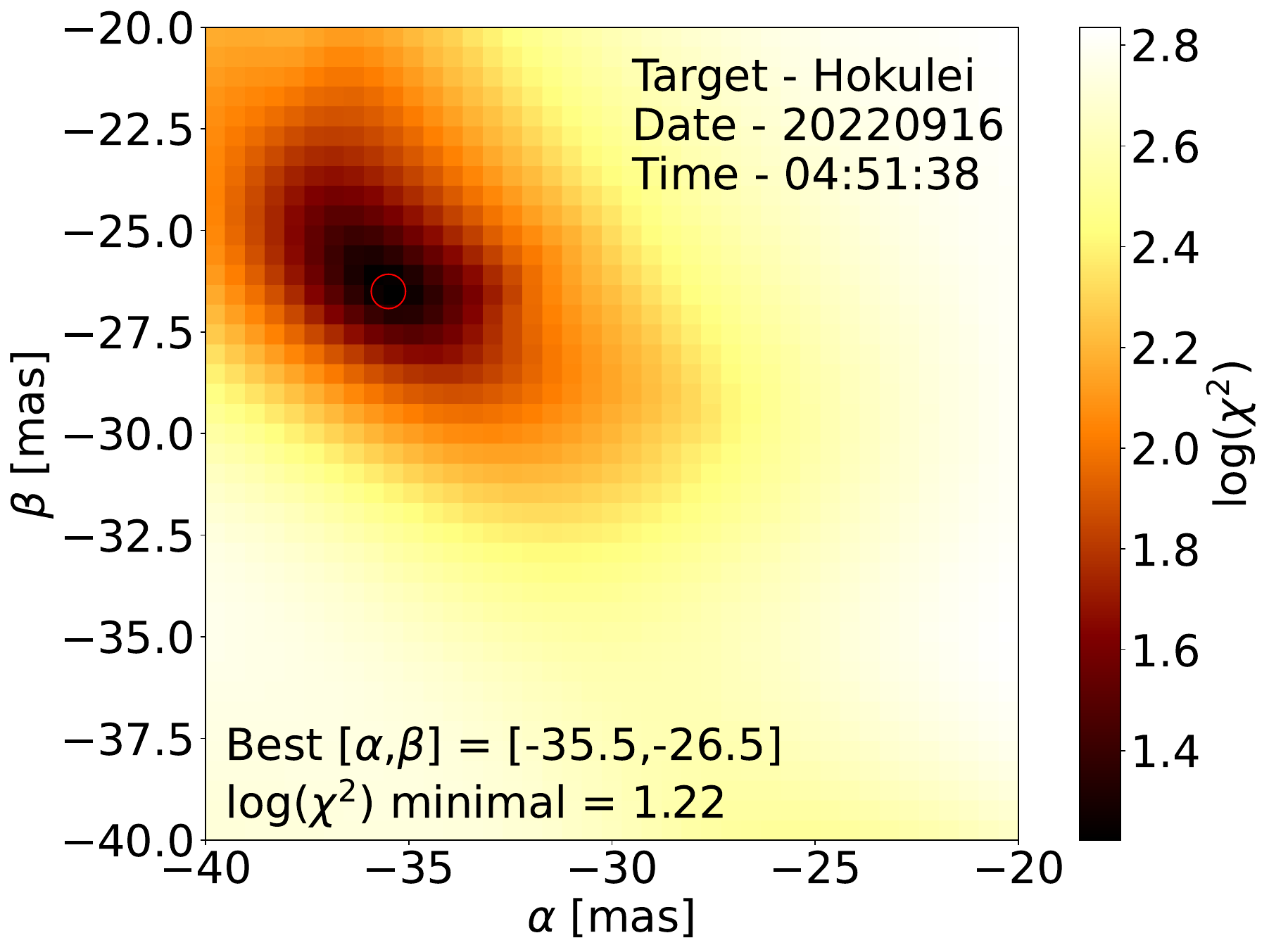} &  \includegraphics[width=0.32\linewidth]{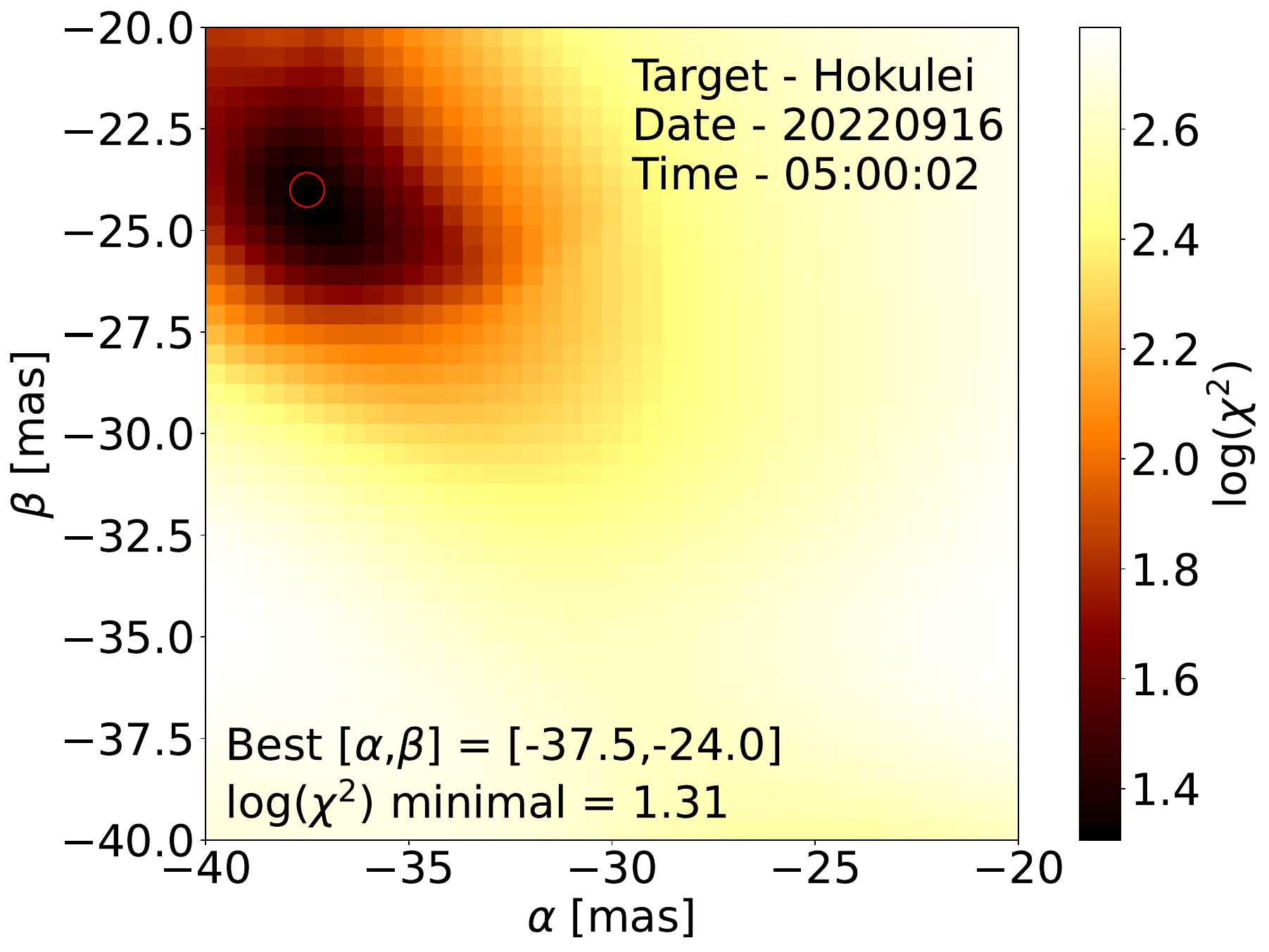} \\
    \end{tabular}
    \caption{Estimation of the Hokulei binary companion position delivered by \revision{$\chi^2$} maps, for 3 cubes of data during our sequence. Time increases from left to right. We can notice that the position estimate evolves with time due to the field rotation.}
    \label{fig:Likelihood}
\end{figure*}


\subsubsection{Calibration of the field rotation}

On September 16th 2020, we acquired 19~cubes of 500~frames each on Hokulei. The exposure time was $30$~ms and the EM gain was 300. For calibration purposes, we acquired 20~cubes of 500~frames each on $\rho$~Persei. The exposure time was $40$~ms and the EM gain was 300. Following the same procedure explained in Sect.~\ref{sec:Vega} we extracted the 28~baseline complex coherences from the 8~\rev{subaperture}s, and computed the 56~CPs. We show on Fig.~\ref{fig:CP_Capella} the signal of the 56 CPs (in blue on the graphs) extracted from one of the Hokulei data cubes, and calibrated with the $\rho$~Persei calibrator measurements. We can differentiate two categories in these CP plots. 1- the triangles which do not detect the companion. These triangles are formed by baselines whose on-sky projected spatial frequency do not overlap with the spatial frequency of the binary. It is the case, for example, of triangle \#3 where we can see that the CP is $0^\circ$ (similarly to the Keho`oea CP measurements) over the spectral bandwidth. 2- the triangles which detect the companion. In this case, the on-sky projected spatial frequency overlaps with the binary spatial frequency, creating a \rev{nonzero} phase of the object visibility. It is the case of, for example, triangle \#49. 

We use Eq.~\ref{eq:bin-vis} to model CP measurements for different flux ratios and angular separations between the two components of the binary. We then compare the generated models to the CPs obtained on-sky using Eqs.~\ref{eq:chi2} and~\ref{eq:likelihood}. This allows us to build $\chi^2$ and likelihood maps - one per spectral channel - to estimate the best ($\alpha,\beta$) position of the binary companion. We show \revision{an example of $\chi^2$} maps on Fig.~\ref{fig:Likelihood}, providing an estimate of the companion position. We display three~maps compiled from three different cubes of 500 images each (cubes number \revision{2, 10 and 19}). The best CP fit (computed from cube number 2) is also displayed on the graphs of Fig.~\ref{fig:CP_Capella}, where we can see that the model is very similar to the on-sky measurements. 

One interesting thing to highlight is the evolution of the estimated position on the likelihood maps. We can see that it starts at a position around $(-34,-27)$~mas, and ends around $(-38,-24)$~mas. This is due to field rotation, since the imaging mode on SCExAO is with a fixed pupil. This variation should be calibrated by the derotation step of the data reduction. In order to retrieve the absolute on-sky position angle of the companion, we have to subtract offset angles due to the fixed pupil imaging mode, and different optics in the AO188 and SCExAO instruments. Using the Hokulei data sets provides an opportunity to validate this calibration. First, we compute the parallactic angle, which is corrected by an image rotator in AO188 to keep the pupil in fixed position. To do so, we use the BATs package, allowing \rev{one} to compute the parallactic angle (noted PAD, see Table~\ref{table_capella_0916}) from the pointing status of the Subaru Telescope\rev{:} the Azimuth and Elevation coordinates of the target (\rev{Az},\rev{El}). We then found the static offset between the FIRST baselines and the pupil by comparing \correction{the averaged PA estimate from the September 16th 2020 data (after subtracting the parallactic angle)} to the theoretical position angle of the binary, given in Table~\ref{table:results}. The static offset computed is $138.2^\circ$. We use the following equation to derotate the FIRST estimated position angle $PA_{FIRST}$ using the static offset and the parallactic angle\rev{:} 
\begin{equation}
    PA_{onsky} = -1\times(PA_{FIRST}-(138.2+PAD))\rev{.}
    \label{eq:PAcorrection}
\end{equation}
We apply this equation to every FIRST estimated position angle \correction{(see Table~\ref{table_capella_0916})} in order to get a de-rotated estimate. \rev{We n}ote that this offset should be the same to apply for all future data sets, assuming the alignment of the pupil stays the same.

\subsubsection{Errors on the estimate}

We now discuss how we compute error bars for the estimated separation and position angle. Each cube $i$ yields an estimated 2D position ($\alpha_i,\beta_i$) of the binary companion, which is then converted into separation and position angle ($Sep_i$, $PA_i$); $i$ being here the cube identification. The final random error is computed as the standard deviation of the cubes averaged estimates divided by $\sqrt{(N-1)}$, with N the number of cubes. These errors are the ones presented in Table~\ref{table:results}. \revision{We also need to take into account a systematic error on the plate scale. It is proportional to the baseline length divided by the wavelength. We estimate the wavelength uncertainty to be 0.5nm, so around $0.1\%$. However, most of the systematic error on the separation comes from the baseline length.} We estimated the \rev{subaperture} diameter to be about $1.04$~meters from the instrument design, but we have no independent way to verify this value. Reducing this systematic error will come, in the future, with calibrating the baseline lengths using well known binaries. We discuss the amplitude of the systematic errors in our data analysis.


\begin{table*}
		\centering
	\caption{Summary of the results: expected (from~\cite{torres2015capella}) and estimated position of the Hokulei binary companion on both days of observation from the collected data. The errors presented here are computed as the standard deviation of the averaged estimates provided by each cubes, hence the standard deviation of the cube estimates divided by $\sqrt{(N-1)}$ (with $N$ the number of cubes).}
	\label{table:results}
	\begin{tabular}{cccccc}
		\hline \hline
		 Date & Expected     &     Expected      & Estimated    &     Estimated    & \revision{Data} \\
           & separation   &  position angle   & separation   &  position angle  & \revision{covariance}     \\
		\hline 
		Sept. 16th 2020  &  $45.1$ mas  &  $274.7$ degrees & $44.3\pm0.1$ mas &  $274.7\pm0.1$ degrees & \revision{$-0.02 \text{ mas}^2$} \\ 
		Sept. 17th 2020  &  $46.0$ mas  &  $270.7$ degrees & $46.4\pm0.1$ mas &  $270.6\pm0.2$ degrees & \revision{$ 0.32 \text{ mas}^2$} \\
		\hline \hline
		\end{tabular}
\end{table*}
\begin{figure*}[h!]
    \centering
    \includegraphics[width=\linewidth]{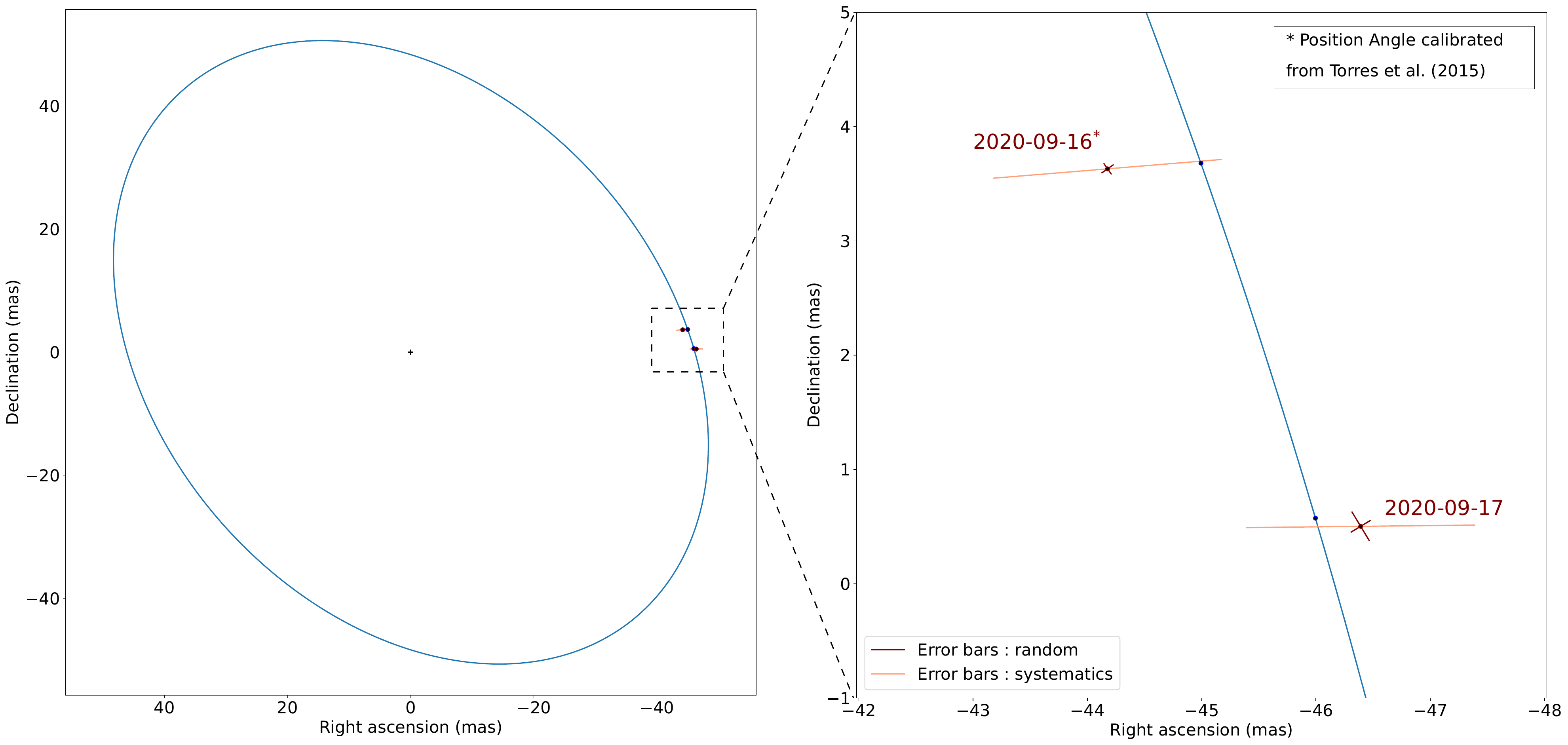}
    \caption{\correction{Orbit of Hokulei Ab binary component. In blue and brown are respectively the expected and estimated positions of the Ab component on both observation days. Brown colored error bars only take into account the random errors. Salmon colored error bars are a quadratic sum of random and systematic errors.}}
    \label{fig:Capella_pos}
\end{figure*}{}

\subsubsection{\correction{Relative measurements of Hokulei components}}

For the 16th of September 2020 data, we find an estimated separation of $44.3\pm0.1$~mas, with a PA of $274.7^\circ\pm0.1^\circ$ \correction{(after applying Eq.~\ref{eq:PAcorrection} to each $PA_{FIRST}$ estimates, as seen in Table~\ref{table_capella_0916}). When comparing with the $45.1$~mas expected separation, we see that our measurement is off by about $1$~mas. This would correspond to a systematic error of about $3\%$ on the \rev{subaperture} size. We show on Fig.~\ref{fig:Capella_pos} our results with and without considering the separation systematic error of $1$~mas for the error bars.}
On September 17th, we acquired 19 cubes of 500~frames each on Hokulei. The exposure time was $10$~ms and the EM gain was 300. For calibration purposes, we acquired 20~cubes of 100 frames each on the SCExAO calibration source right after closing the telescope. The exposure time was $10$~ms with no EM gain. Using the same procedure as for the 16th September 2020, we compute the CP and perform the model fitting. We find an estimated separation of $46.4\pm0.1$~mas with a PA of $270.6\pm0.2^\circ$, from the data cubes analysis. As we can see in the Table~\ref{table:results}, this compares well with the expected separation and PA, respectively of $46$~mas and $270.7^\circ$, taking into account the systematic error. 

We plot on Fig.~\ref{fig:Capella_pos} the orbit of Hokulei Ab, and show the expected and estimated position of the Ab binary component for both dates of observation. \revision{Two error bars are drawn:} error bars that only take into account the random errors, and error bars that \revision{only take into account} systematic errors. 
\revision{The random error bars are drawn along the principal axes of the error ellipse derived from the covariance matrix (see Table~\ref{table:results}). The systematics error bars represents the 1~mas error over the separation, and are oriented \rev{toward} the central star. }We can see how well our estimate of the companion position compares to the expected position \correction{when taking into account the systematic errors}. We can also see that even though we used the SCExAO calibration source to calibrate the CP obtained on Hokulei on September 17th and not a reference star, the estimation of the position compares well with the expected position. Most of the CP bias is well calibrated by the sole SCExAO internal source. Our tests did not allow \rev{us} to provide an absolute estimation of the Ab binary component, since we used \citep{torres2015capella} to de-rotate the data from September 16th 2020, and used this reference for the September 17th 2020 data reduction. However, observing Hokulei at two different epochs allows us to compute the differential PA of the Ab binary component. \citep{torres2015capella} informs us that the differential PA between the two epochs is $4.0^{\circ}$. FIRST provides a differential PA of $3.9^{\circ}\pm0.3^{\circ}$, which compares well with the expected value. In the course of about 24~hours, we were able to track the movement of the companion along its orbit.

\section{Conclusion}

We presented the integration, characterization and on-sky demonstration of the FIRST spectro-interferometer on the \mbox{$8.2$-m} Subaru Telescope. FIRST interferes simultaneously multiple parts of the telescope pupil to extract information from astronomical objects, with a sensitivity to structure that extends down to at least $10$~mas and a spectral resolution of about $300$ at $700$~nm. FIRST's optimal field of view is up to $40$~mas, over which the fiber injection losses are less than $5~\%$. Using 8~samples of the Subaru Telescope pupil, each with a projected diameter of about $1$~m, we demonstrated FIRST's ability to provide CP measurements both on a point source - Keho‘oea - and on a binary star\rev{:} Hokulei. \revision{The Keho‘oea observation allowed to highlight several improvements of the instrument compared to the previous setup installed on the 3-meter Lick Telescope. The instrument now shows better efficiency in terms of acquisition procedure and better stability allowing the exploitation of all data and measurements thus increasing the overall quality of the results. The Keho‘oea data showed an average accuracy of $0.6^\circ$ from the CP measurements, and statistical error of $0.15^\circ$ at best on uncalibrated data. We also evaluated FIRST detection capabilities. Analysis showed ability to sense companions separated from their host star with a separation as little as a quarter of the telescope theoretical diffraction limit. Moreover, we assessed that the instrument can enable detection with a contrast down to 0.02 at $\lambda/D$ separation. The observation of Hokulei} demonstrated FIRST's capability to detect and track the position of the binary companion along its orbit using observations about 24~hours apart. The best random error on the separation and PA were respectively $0.1$~mas and $0.1^\circ$. \correction{Systematic errors on the separation, due to an uncertainty on the \rev{subaperture} projected diameter, were estimated to be about 1~mas. Acquiring more data on well known binary systems will be essential to calibrate systematic errors.}
With these new capabilities, and the demonstration of the instrument to perform measurements below the diffraction limit of the telescope, we will be able to fully commission FIRST at the Subaru telescope and offer it to the scientific community for open use.

The success of demonstrating/commissioning such instrument on SCExAO is an important stepping stone for future interferometric instrumentation on extremely large telescopes. \correction{Their large collecting area and small diffraction limit (about 4~mas in the optical wavelengths), coupled with spectral dispersion capabilities, would offer unique high contrast and high resolution capabilities with high sensitivity for Earth-like exoplanet research and characterization. }

\begin{acknowledgements}
The development of FIRST was supported by Centre National de la Recherche Scientifique CNRS (Grant ERC LITHIUM - STG - 639248).
The development of SCExAO was supported by the Japan Society for the Promotion of Science (Grant-in-Aid for Research \#23340051, \#26220704, \#23103002, \#19H00703 \& \#19H00695), the Astrobiology Center of the National Institutes of Natural Sciences, Japan, the Mt Cuba Foundation and the director’s contingency fund at Subaru Telescope. 
N.C. acknowledges funding from the European Research Council (ERC) under the European Union’s Horizon 2020 research and innovation program (grant agreement CoG - 683029).
K.A. acknowledges funding from the Heising-Simons foundation.
V.D. and N.S. acknowledge support from NASA funding (Grant \#80NSSC19K0336).
N.S. acknowledges support from the PSL Iris-OCAV project.
M.T. is supported by JSPS KAKENHI grant Nos.18H05442, 15H02063, and 22000005.
S.V. would like to express gratitude to Kumu Leilehua Yuen, Leinani Lozi and Amy Durham for the enlightening, inspiring and helpful discussions about Hawai`i and Native Hawaiian culture, and how to incorporate it in this paper. The authors wish to recognize and acknowledge the very significant cultural role and reverence that the summit of Mauna Kea has always had within indigenous Hawaiian communities, and are most fortunate to have the opportunity to conduct observations from this mountain.
\end{acknowledgements}

\bibliography{main} 
\bibliographystyle{bibtex/aa}

\appendix
\numberwithin{equation}{section}
\renewcommand{\theequation}{\thesection\arabic{equation}}
\numberwithin{figure}{section}
\renewcommand{\thefigure}{\thesection\arabic{figure}}

\section{Reconstruction of the baseline complex coherences}
\label{Appendix-datared}

Considering N fibers providing N interfering beams, the fringe intensity can be written and developed, for each wavelength, as: 
\begin{align}
I(\Vx)&= \GP{\sum_{n<N}A_nE_n(\Vx)}^2 \nonumber\\
&= \sum_{n<N}A_n^2E_n^2(\Vx) \nonumber\\ 
&+ 2 \text{ Re}\GC{\sum_{n<n'<N} A_nE_n(\Vx)A_{n'}E_{n'}(\Vx)\text{e}^{i(2\pi f_{nn'}\Vx+\Delta\Phi_{nn'})}} \nonumber\\
&= \sum_{n<N}A_n^2E_n^2(\Vx) \label{eq2} \\ 
&+ 2 \sum_{n<n'<N} A_nE_n(\Vx)A_{n'}E_{n'}(\Vx)\text{cos}\GP{2\pi f_{nn'}\Vx+\Delta\Phi_{nn'}}\nonumber
\end{align}
with $\Vx$ the spatial variable (pixel index), $E_n$ the normalized envelope of each beam and $f_{nn'}$ the $nn'$ baseline frequency. By taking into account the object's complex visibility and \rev{rearranging} Eq.~\ref{eq2}, it is possible to obtain a linear relationship between the interferogram $I(\Vx)$ and the complex coherence (Eq.~\ref{eq:mu}):
\begin{align}
I(\Vx) = & \mu_0E_g(\Vx)\label{Eq3} \\ & +\sum_{n<n'<N}\mathcal{R}\{\ \mu_{nn'}\}C_{nn'}(\Vx) \nonumber\\ & +\sum_{n<n'<N}\mathcal{I}\{\ \mu_{nn'}\}S_{nn'}(\Vx)\nonumber
\end{align}
with $\mu_0$ the total flux, $Eg(\Vx)$ the normalized global envelope function and
\begin{align}
\left\{  
\begin{array}{r@{}l}\displaystyle
&C_{nn'}(\Vx)= 2E_n(\Vx)E_{n'}(\Vx)\cos(2\pi f_{nn'}\Vx)\nonumber\\
&S_{nn'}(\Vx)=-2E_n(\Vx)E_{n'}(\Vx)\sin(2\pi f_{nn'}\Vx)\nonumber
\end{array}\right. 
\end{align}
$\{ E_g(\Vx) \text{;} C_{nn'}(\Vx) \text{;} S_{nn'}(\Vx)  \}$ form a base on which the $I(\Vx)$ interferogram can be decomposed. This base depends on the instrument setup. Let us define $x_k$ the pixel index with $k \in \{1,...,n_p\} $ with $n_p$ the number of pixels. We also concatenate indexes $nn'$ into $\{1,...,n_B\}$ with $n_B$ the number of baseline pairs.  Finally the unknowns ($\mu_{n_B}$) can then be regrouped in a vector \textbf{P} and Eq.~(\ref{Eq3}) can be written as the matrix product:
\begin{align}
\text{\textbf{I}}=
\begin{bmatrix}
	I_{x_1} \\
	\vdots \\
	I_{x_{np}} 
\end{bmatrix}
= \mathrm{V2PM} \cdot
\begin{bmatrix}
\mathcal{R}\{\ \mu_{1} \} \\
\vdots \\
\mathcal{R}\{\ \mu_{n_B} \} \\
\mathcal{I}\{\ \mu_{1} \} \\
\vdots \\
\mathcal{I}\{\ \mu_{n_B} \} \\
\mu_0
\end{bmatrix}
=\mathrm{V2PM}\cdot\text{\textbf{P}}
\end{align}
with V2PM =
\begin{align}
	\begin{bmatrix}\label{v2pm}
	C_1(x_1) & \dots & C_{n_B}(x_1)  & S_1(x_1) & \dots & S_{n_B}(x_1) & E_g(x_1) \\
	\vdots &  & \vdots & \vdots &  & \vdots &   \vdots \\
	C_1(x_{n_p}) & \dots & C_{n_B}(x_{n_p})  & S_1(x_{n_p}) & \dots & S_{n_B}(x_{n_p}) & E_g(x_{n_p})
	\end{bmatrix}\nonumber
\end{align}

According to the formalism introduced by~\cite{millour2004data}, the V2PM is the Visibility-To-Pixel matrix with a size of $(2n_B+1)\times n_p$. This matrix is rectangular hence it cannot be inverted. However, because the recombination on FIRST is \rev{nonredundant}, each baseline frequency is unique hence the V2PM modes are orthogonal one to another. In this case, we can compute V2PM$^\dag$, the partial generalized inverse of V2PM matrix computed from the Singular Value Decomposition of V2PM. We can then compute the estimates $\EST{\textbf{P}}$ according to:
\begin{equation}\label{projection}
	\EST{\textbf{P}}=\mathrm{V2PM}^\dag\cdot\text{\textbf{I}},
\end{equation}

which allow\rev{ed us)} to reconstruct the baselines complex coherences.

\section{Hawaiian naming of stars}
\label{Appendix-hawaiian-names}
The following knowledge was gathered from Kumu Leilehua Yuen, a cultural practitioner and traditional knowledge holder on Hawai`i island and from~\cite{johnson1975inoa}. The intention of sharing this knowledge is to grow our understanding of these main targets and astronomy as a whole through a lense of heritage and knowledge specific to where these observations are made, Hawai`i. In Native Hawaiian culture~\footnote{A distinction is made between the indigenous Native Hawaiian culture and the multicultural “local” practices which have evolved since the in-migration of peoples from around the world.}, celestial objects can change in name depending on their position in the sky and their use. For example, Keho`oea is also named  Kah\={o}`eoa, Kaho`ea or Keoe. The name Keho`oea appears in chant fourteen (Ka Wa Umikumamaha) of the Kumulipo, line 1859~\citep{beckwith2000kumulipo}. The Kumulipo is a genealogical chant and the most well known origin story of Native Hawaiian culture. Like other important cultural knowledge, the Kumulipo was transmitted orally for generations by highly trained specialists. It was documented in writing during the time of King Kal\={a}kaua. One of the most important translations into English was by Hawai`i’s last monarch, Queen Lili`uokalani during her unjust imprisonment within her own palace and the illegal occupation of the Kingdom of Hawai`i during a coup by American businessmen.  \\
In literal translation from `\={o}lelo hawaii (\rev{H}awaiian language), Hokulei is a compound word: “h\={o}k\={u}” / “star” and “lei” / “garland” or “to rise like a cloud.” Some practitioners translate the name as “star garland.” Yuen was taught the translation "star/constellation that rises like a cloud,” in reference to the fact that Hokulei is the brightest star of its constellation, Aurigae.


\section{FIRST raw data}

We present in the following Tables the output of the FIRST data reduction, for every cube taken during the two epochs.
\label{Appendix-onsky-data}
\begin{table*}
		\centering
	\caption{Table of CP fitting results Sept $16^{th}$ 2020. $\hat{\alpha}$: Estimated $\alpha$ - $\hat{\beta}$: Estimated $\beta$ - Sep\rev{:} Estimated Separation - $PA_{FIRST}$: Estimated Position Angle - (Az,El): Azimuth and Elevation coordinates of the telescope - PAD\rev{:} Parallactic Angle -  $PA_{on-sky}$\rev{:} Estimated on-sky Position Angle after de-rotation.}
	\begin{tabular}{cccccccc}
		\hline \hline
		Time (HST)           & $\hat{\alpha}$ (mas)         & $\hat{\beta}$ (mas)        & Sep (mas)  & $PA_{FIRST}$ ($^{\circ}$)    &     (Az, El) ($^{\circ}$)     &   PAD ($^{\circ}$) & $PA_{on-sky}$ ($^{\circ}$)\\
		\hline \hline
				         \multicolumn{8}{c}{\textbf{2020-09-16}} \\
		\hline
		04:43:03  &  $-33.5\pm1.1$  & $-28.5\pm0.7$  & $44.0\pm1.3$   & $-139.6\pm0.9$ &   (205.63,59.824) &   -3.13031  & 274.4697 \\ 
		04:44:00  &  $-34.0\pm0.7$  & $-28.0\pm0.5$  & $44.0\pm0.8$   & $-140.5\pm0.6$ &   (205.36,59.920) &   -3.53590  & 274.9641  \\
		04:44:58  &  $-34.0\pm0.6$  & $-28.0\pm0.4$  & $44.0\pm0.7$   & $-140.5\pm0.5$ &   (205.09,60.017) &   -3.94199  & 274.5580 \\ 
		04:45:54  &  $-34.0\pm0.6$  & $-28.0\pm0.4$  & $44.0\pm0.7$   & $-140.5\pm0.5$ &   (204.82,60.110) &   -4.34559  & 274.1544 \\
		04:46:52  &  $-34.5\pm0.7$  & $-27.5\pm0.4$  & $44.1\pm0.8$   & $-141.4\pm0.6$ &   (204.54,60.204) &   -4.76260  & 274.6374 \\
		04:47:49  &  $-34.5\pm0.6$  & $-27.5\pm0.4$  & $44.1\pm0.7$   & $-141.4\pm0.5$ &   (204.26,60.298) &   -5.17950  & 274.2205 \\
		04:48:46  &  $-35.0\pm0.7$  & $-27.0\pm0.6$  & $44.2\pm0.9$   & $-142.4\pm0.6$ &   (203.98,60.389) &   -5.59457  & 274.8054\\
		04:49:43  &  $-35.0\pm0.4$  & $-27.0\pm0.3$  & $44.2\pm0.5$   & $-142.4\pm0.3$ &   (203.70,60.480) &   -6.00952  & 274.3905 \\
		04:50:40  &  $-35.5\pm0.5$  & $-26.5\pm0.4$  & $44.3\pm0.6$   & $-143.3\pm0.4$ &   (203.42,60.570) &   -6.42379  & 274.8762 \\
		04:51:38  &  $-35.5\pm0.5$  & $-26.5\pm0.3$  & $44.3\pm0.6$   & $-143.3\pm0.4$ &   (203.13,60.660) &   -6.85092  & 274.4491 \\
		04:52:35  &  $-35.5\pm0.5$  & $-26.5\pm0.3$  & $44.3\pm0.6$   & $-143.3\pm0.4$ &   (202.85,60.747) &   -7.26326  & 274.0367 \\
		04:53:31  &  $-36.0\pm0.4$  & $-26.0\pm0.3$  & $44.4\pm0.5$   & $-144.2\pm0.3$ &   (202.56,60.832) &   -7.68742  & 274.5126\\
		04:54:29  &  $-36.5\pm0.4$  & $-25.5\pm0.3$  & $44.5\pm0.5$   & $-145.1\pm0.3$ &   (202.27,60.919) &   -8.11253  & 274.9875 \\
		04:55:26  &  $-36.5\pm0.4$  & $-25.0\pm0.3$  & $44.5\pm0.5$   & $-145.1\pm0.3$ &   (201.97,61.003) &   -8.54898  & 274.5510\\
		04:56:24  &  $-36.5\pm0.4$  & $-25.5\pm0.3$  & $44.5\pm0.5$   & $-145.1\pm0.3$ &   (201.67,61.088) &   -8.98584  & 274.1142 \\
		04:57:18  &  $-37.0\pm0.5$  & $-25.0\pm0.3$  & $44.7\pm0.6$   & $-146.0\pm0.4$ &   (201.39,61.166) &   -9.39276  & 274.6072 \\
		04:58:12  &  $-37.5\pm0.5$  & $-24.5\pm0.3$  & $44.8\pm0.6$   & $-146.8\pm0.4$ &   (201.11,61.243) &   -9.79904  & 275.0009\\
		04:59:07  &  $-37.5\pm0.5$  & $-24.5\pm0.3$  & $44.8\pm0.6$   & $-146.8\pm0.4$ &   (200.82,61.320) &   -10.2183  & 274.5817 \\
		05:00:02  &  $-37.5\pm0.5$  & $-24.0\pm0.4$  & $44.5\pm0.7$   & $-147.5\pm0.3$ &   (200.52,61.397) &   -10.6506  & 274.8494 \\
		\hline \hline
		\end{tabular}
	\label{table_capella_0916}
\end{table*}

\begin{table*}
    \centering
    \caption{Table of CP fitting results Sept $17^{th}$ 2020. $\hat{\alpha}$: Estimated $\alpha$ - $\hat{\beta}$: Estimated $\beta$ - Sep\rev{:} Estimated Separation - $PA_{FIRST}$: Estimated Position Angle - (Az,El): Azimuth and Elevation coordinates of the telescope - PAD\rev{:} Parallactic Angle -  $PA_{on-sky}$\rev{:} Estimated on-sky Position Angle after de-rotation.} 
	\begin{tabular}{cccccccc}
		\hline \hline
		Time (HST)  & $\hat{\alpha}$ (mas)  & $\hat{\beta}$ (mas)  & Sep (mas)  & $PA_{FIRST}$ ($^{\circ}$) &  (Az, El) ($^{\circ}$)     &   PAD ($^{\circ}$) & $PA_{on-sky}$ ($^{\circ}$) \\
		\hline \hline
						         \multicolumn{8}{c}{\textbf{2020-09-17}} \\
		\hline
		05:07:55  &  $-39.5\pm2.0$  & $-23.0\pm1.2$  & $45.7\pm2.3$  &  $-149.8\pm1.3$ & (196.59,62.284) & -16.2458 &  271.7542 \\
		05:08:18  &  $-39.5\pm1.4$  & $-23.0\pm1.3$  & $45.7\pm2.1$  &  $-149.8\pm0.9$ & (196.46,62.310) & -16.4291 &  271.5709 \\
		05:08:40  &  $-39.5\pm1.5$  & $-23.0\pm1.3$  & $45.7\pm2.2$  &  $-149.8\pm0.9$ & (196.33,62.334) & -16.6115 &  271.3885 \\
		05:09:02  &  $-39.5\pm1.4$  & $-23.0\pm1.3$  & $45.7\pm2.1$  &  $-149.8\pm0.9$ & (196.21,62.359) & -16.7810 &  271.2190 \\
		05:09:26  &  $-40.0\pm1.4$  & $-22.5\pm1.1$  & $45.9\pm1.9$  &  $-150.6\pm0.9$ & (196.07,62.385) & -16.9775 &  271.8225 \\
		05:09:49  &  $-40.0\pm1.2$  & $-22.5\pm1.0$  & $45.9\pm1.7$  &  $-150.6\pm0.7$ & (195.93,62.410) & -17.1736 &  271.6264 \\
		05:10:12  &  $-40.0\pm1.4$  & $-22.5\pm1.2$  & $45.9\pm2.0$  &  $-150.6\pm0.9$ & (195.80,62.434) & -17.3560 &  271.4440 \\
		05:10:35  &  $-40.0\pm1.3$  & $-22.5\pm1.2$  & $45.9\pm2.0$  &  $-150.6\pm0.8$ & (195.66,62.459) & -17.5520 &  271.248  \\
		05:10:58  &  $-40.5\pm1.2$  & $-22.0\pm1.0$  & $46.1\pm1.7$  &  $-151.5\pm0.7$ & (195.53,62.483) & -17.7343 &  271.9657 \\
		05:11:22  &  $-40.5\pm1.5$  & $-22.0\pm1.2$  & $46.1\pm2.1$  &  $-151.5\pm0.9$ & (195.39,62.508) & -17.9303 &  271.7697 \\
		05:15:56  &  $-40.5\pm0.6$  & $-22.0\pm0.6$  & $46.1\pm1.0$  &  $-151.5\pm0.4$ & (193.76,62.779) & -20.2034 &  269.5966 \\
		05:16:39  &  $-41.5\pm0.7$  & $-20.5\pm0.7$  & $46.3\pm1.2$  &  $-153.7\pm0.4$ & (193.50,62.819) & -20.5645 &  271.3355 \\
		05:17:21  &  $-42.0\pm0.6$  & $-20.0\pm0.7$  & $46.5\pm1.1$  &  $-154.5\pm0.3$ & (193.25,62.857) & -20.9116 &  271.7884 \\
		05:18:04  &  $-41.5\pm0.7$  & $-20.5\pm0.7$  & $46.3\pm1.2$  &  $-153.7\pm0.4$ & (192.99,62.896) & -21.2723 &  270.6277 \\
		05:18:47  &  $-42.0\pm0.5$  & $-20.0\pm0.7$  & $46.5\pm1.1$  &  $-154.5\pm0.3$ & (192.73,62.933) & -21.6323 &  271.0677 \\
		05:19:29  &  $-42.0\pm0.5$  & $-20.0\pm0.5$  & $46.5\pm0.9$  &  $-154.5\pm0.3$ & (192.47,62.969) & -21.9919 &  270.7081 \\
		05:20:11  &  $-42.0\pm0.5$  & $-20.0\pm0.5$  & $46.5\pm0.9$  &  $-154.5\pm0.3$ & (192.22,63.005) & -22.3380 &  270.3620 \\
		05:20:54  &  $-42.0\pm0.4$  & $-19.5\pm0.5$  & $46.3\pm0.8$  &  $-155.1\pm0.2$ & (191.95,63.040) & -22.7106 &  270.5894 \\
		05:21:36  &  $-42.0\pm0.6$  & $-19.5\pm0.5$  & $46.1\pm0.9$  &  $-155.7\pm0.3$ & (191.69,63.074) & -23.0695 &  270.8305 \\
		05:22:19  &  $-42.5\pm0.4$  & $-19.0\pm0.5$  & $46.6\pm0.8$  &  $-155.9\pm0.2$ & (191.43,63.108) & -23.4282 &  270.6718 \\
		05:30:33  &  $-43.5\pm0.3$  & $-17.0\pm0.3$  & $46.7\pm0.6$  &  $-158.7\pm0.1$ & (188.33,63.441) & -27.6822 &  269.2178 \\
		05:31:14  &  $-44.0\pm0.8$  & $-16.5\pm0.8$  & $47.0\pm1.6$  &  $-159.4\pm0.3$ & (188.07,63.464) & -28.0375 &  269.5625 \\
		05:32:37  &  $-44.0\pm1.8$  & $-16.0\pm1.3$  & $46.8\pm2.9$  &  $-160.0\pm0.8$ & (187.54,63.508) & -28.7611 &  269.4389 \\
		05:33:20  &  $-44.0\pm0.3$  & $-16.0\pm0.3$  & $46.8\pm0.6$  &  $-160.0\pm0.1$ & (187.26,63.530) & -29.1431 &  269.0569 \\
		05:34:03  &  $-44.5\pm0.3$  & $-15.5\pm0.2$  & $47.1\pm0.5$  &  $-160.8\pm0.1$ & (186.99,63.551) & -29.5113 &  269.4887 \\
		05:34:46  &  $-44.5\pm0.3$  & $-15.0\pm0.3$  & $47.1\pm0.6$  &  $-160.8\pm0.1$ & (186.71,63.571) & -29.8929 &  269.1071 \\
		05:35:29  &  $-45.0\pm0.3$  & $-15.0\pm0.3$  & $47.4\pm0.6$  &  $-161.6\pm0.1$ & (186.43,63.591) & -30.2744 &  269.5256 \\
		05:36:12  &  $-45.0\pm0.3$  & $-14.5\pm0.3$  & $47.3\pm0.6$  &  $-162.1\pm0.1$ & (186.15,63.609) & -30.6555 &  269.6445 \\
		05:36:53  &  $-45.0\pm0.2$  & $-14.0\pm0.3$  & $47.1\pm0.6$  &  $-162.7\pm0.1$ & (185.89,63.626) & -31.0094 &  269.8906 \\
		\hline \hline      

	\end{tabular}
	\label{table_capella_0917}
\end{table*}

\end{document}